\def\hst{{\sl HST}}
\def\nicmos{NICMOS}
\def\gem{{\sl Gemini}}
\def\gni{GNIRS}
\def\nif{NIFS}
\def\spitzer{{\sl Spitzer}}
\def\irac{IRAC}

\documentclass[12pt,preprint]{aastex}
\usepackage{appendix}
\usepackage{epsf}
\usepackage{color}
\usepackage{graphicx}
\usepackage{graphics}
\usepackage{epsfig}
\usepackage{longtable}
\usepackage{multirow}
\begin{document}
\title{Origins of Massive Field Stars in the Galactic Center: a
  Spectroscopic Study}
\author{Hui Dong$^{1}$, Jon Mauerhan$^{2,3}$, Mark R. Morris$^4$,
  Q. Daniel Wang$^5$, Angela Cotera$^6$}

\affil{$^1$ National Optical Astronomy Observatory,
Tucson, AZ, 85719}
\affil{$^2$ Department of Physics, University of California, Berkeley,
  CA, 94720}
\affil{$^3$ Steward Observatory, University of Arizona, Tucson, AZ, 85719}
\affil{$^4$Department of Physics and Astronomy, University of California, Los
Angeles, CA, 90095}
\affil{$^5$ Department of Astronomy, University of Massachusetts,
Amherst, MA, 01003}
\affil{$^6$ SETI Institute, Mountain View, CA, 94043}
\affil{E-mail: hdong@noao.edu}

\begin{abstract}
Outside of the known star clusters in the Galactic Center, a large number of evolved massive stars have been detected; but their origins remain uncertain. We present a spectroscopic study of eight such stars, based on new \gem\ GNIRS and NIFS near-infrared observations. This work has led to the discovery of a new O If$^+$ star. We compare the reddening-corrected $J$$-$$K$ vs $K$ diagram for our stars with the massive ones in the Arches cluster and use stellar evolutionary tracks to constrain their ages and masses. The radial velocities of both the stars and their nearby \ion{H}{2} regions are also reported. All of the stars are blueshifted relative to the Arches cluster by $>$ 50 km/s. We find that our source P35 has a velocity consistent with that of the surrounding molecular gas. The velocity gradient of nearby ionized gas along the \gem\ GNIRS long slit, relative to P35 and the adjacent -30$-$0 km/s molecular cloud, can best be explained by a pressure-driven flow model. Thus, P35 most likely formed in situ. Three more of our stars have radial velocities different from their adjacent molecular gas, indicating that they are interlopers. The four stars closest to the Arches cluster have similar spectra, ages and masses to known cluster members, suggesting that they were likely ejected from the cluster via three-body interactions. Therefore, we find that the relatively isolated stars partly formed in situ and partly were ejected from the known star clusters in the Galactic Center. 
\end{abstract}

\section{Introduction}\label{s:intro}
As the closest galactic nucleus, the Galactic Center (GC), containing its
supermassive black hole (SMBH), Sgr A*~\citep[$\sim$4$\times$10$^6$ M$_{\odot}$,][]{ghe08,gil09}, provides a unique laboratory for studies 
of the interplay of stars with the interstellar medium (ISM) in galactic circumnuclear
environments. Because of the
proximity of the GC ($\sim$8.4 kpc,~\citealt{ghe08,gil09}), we are able to reach
resolutions unparalleled in external galaxies (1\arcsec =0.04 pc) and distinguish individual
stars. The central $\sim$300 pc of Galaxy,
known as the central molecular zone (CMZ), is composed of
$\sim$4$\times$10$^7$ M$_{\odot}$ of 
molecular gas, much of which follows a twisted
elliptical ring~\citep{mor96a,mol11}. Several molecular clouds are
collapsing and forming stars at an estimated rate of $\sim$0.03 M$_{\odot}$/yr~\citep{lon13}. This is much different from the
circumnuclear region of M31,
the second closest galactic nucleus, which has small amount of
molecular gas and no sign of current star
formation~\citep[][and references therein]{li09}. Therefore, the GC is the best place to study the mode of star formation in the vicinity of a SMBH in
great detail. A comprehensive study of massive stars in the GC
is important to address several questions: 1) How do molecular
clouds form stars in galactic nuclei? 2)
Are massive clusters preferred in the GC, due to the large
Jeans Mass~\citep{mor96a}? 3) What is the star formation
efficiency of molecular clouds in the GC? What fraction of molecular
clouds is converted into stars, before spiraling into the
potential well of, and thereby feeding, the SMBH? The answer to the
last question can 
help determine the long-term activity of Sgr A*, which thereby
affects 
the evolutionary history of the central regions of the Galaxy.

Over the past two decades, studies on current star formation 
in the GC have been centered on three young and massive star clusters 
within the inner 30 pc of the GC (the Arches, Quintuplet
and Central clusters, $\sim$10$^4$ M$_{\odot}$,
$<$10 Myr old,~\citealt{fig99,fig02,gen03}). These clusters are so
bright and compact that they clearly stand out in ground-based near-infrared
(NIR) surveys, even amidst the strong stellar background
toward the GC. These clusters are
considered to be 
low-mass analogs of super star clusters in external galaxies, harboring hundreds 
of young massive stars whose strong stellar winds shape 
nearby molecular clouds~\citep{lan99,lan01a}. 
The origin of these clusters is not yet well understood. 
Ground-based adaptive optics (AO) observations find that the Arches
and Quintuplet clusters are moving toward Galactic East with velocities of 
172$\pm$15 km/s~\citep{cla12} and 132$\pm$15
km/s~\citep{sto14}. The radial velocities of the Arches and Quintuplet 
cluster are 95$\pm$8 km/s~\citep{fig02} and 102$\pm$2
km/s~\citep{lie09,sto14}.~\citet{sto14} model 
the possible trajectories of the two clusters 
based on their 3D space 
velocities, and conclude that they may have formed in the 
transition zone between the Galactic bar Lindblad x1 and x2 orbits.

In addition to these three well-studied clusters,
ever-increasing efforts have been undertaken to search for 
young stars outside the three known clusters~\citep{cot99,hom03,mik06,mun06,mau10a}. 
Because the three clusters contribute only about half of the Lyman
continuum photons in the GC~\citep{fig04}, we expect that a similar 
number of massive stars should be situated outside the clusters. Indeed,
our recent \hst/NICMOS Paschen-$\alpha$ survey of the GC reveals 180
Paschen-$\alpha$ emitting sources (PESs), $\sim$100 of which are outside
these clusters (\citealt{don12}). These PESs should
be evolved massive stars 
with strong stellar winds. For example, in the first follow-up
spectroscopic observations,~\citet{mau10b} successfully identify one
of our new sources as a new luminous blue variable (LBV)
surrounded by a circular nebula and~\citet{mau10c}
confirm that an additional 19 of our PESs are either OB supergiants or 
Wolf-Rayet (WR) stars. Unlike the PESs in the 
clusters, our `field' PESs are relatively isolated, or are within
small groups. Some of our PESs are coincident with \ion{H}{2} 
regions~\citep{don12}. These PESs could be: 1) previous 
members of the three clusters, ejected by
three-body interactions, binary-supernova explosion, or stripped off
by a very strong tidal force, particularly 
when the clusters passed by Sgr A*;
 2) members of dissolved massive star clusters;
3) members of young and less massive star clusters with only a few massive
stars. The
locations and properties of any `field' massive stars are essential to properly constraining the star
formation history and mode of the GC. 

In this paper, we present new \gem\ spectroscopic observations, and
use them to measure the
radial velocities of the PESs to the Galactic west
($<$10\arcmin , i.e. 24 pc) of the Arches cluster. This region 
includes more than one dozen PESs~\citep{don12}, most of which 
have been identified as OB supergiants, and a few WR stars
~\citep{cot99,mun06,mau10a,mau10c}, as well as the 
Arched filaments and several compact \ion{H}{2} 
regions (H1 to H13)~\citep{mor89,lan01b}. 
This particular region was selected because these PESs lie 
behind the proper motion vector of the Arches cluster and could be its  
deserters, owing either to the regions tidal force or three-body
interaction. Another reason is that~\citet{don12} 
report that several of these PESs are sufficiently luminous that their 
ultraviolet photons are capable of ionizing the nearby \ion{H}{2} 
regions, indicating their possible physical association, 
not merely a line-of-sight coincidence. However, it is
still not clear whether 
these stars are physically associated with the nearby, possibly natal, molecular
clouds, or are massive stars 
only just now encountering these clouds in the course of their
orbits. Here, we want to
use the radial velocities of these PESs, as well as their
extinction-corrected 
colors and magnitudes, to further explore their relationship with the
known star clusters, especially the Arches
cluster, and nearby \ion{H}{2} regions. 

In \S\ref{s:obs}, we
introduce our program objects and report on the 
spectroscopic observations. In~\S\ref{s:analysis}, 
we determine the spectral types of the PESs and compare their  
extinction-corrected colors and magnitudes with the massive 
stars of the Arches cluster and with stellar evolutionary tracks. We 
also measure the radial velocities of the stars and nearby \ion{H}{2} regions. 
We discuss our results in \S\ref{s:dis} and summarize in \S\ref{s:sum}.  

\section{Sample and Spectroscopic Observations}\label{s:obs}
\subsection{Sample Selection}\label{ss:sam}
We have selected OB supergiants and O If$^{+}$ stars\footnote{The
  superscript `+' indicates the existence of \ion{S}{4}  
  (4089\AA\ and 4116\AA ) and \ion{N}{3}/\ion{He}{2} 4686\AA\  \citep{wal71}.} associated with
extended emission in our \hst/\nicmos\ GC Paschen-$\alpha$
image~\citep{wan10,don12}. The spectra of these stars still
 show helium absorption lines~\citep{mar08}, 
indicating that their stellar winds are relatively weak, compared to the WR stars.  
Therefore, the centroids of these absorption lines, as well as several forbidden emission 
lines (see \S\ref{ss:vel_s}), from the photospheres  
could represent the radial velocities of the stars. In contrast, the spectra
of WR stars are dominated by intense
emission lines from expanding stellar winds, which are always too
broad to be used to 
accurately determine the stellar systemic velocities.  

We have obtained the NIR spectra of eight Paschen-$\alpha$ emitting OB
supergiants and O If$^{+}$ stars. The salient parameters
 of these stars are listed in Table~\ref{t:sample}. The star names
 used throughout this paper are
 adopted from~\citet{don12}. 
Seven of these stars are less than 5\arcsec\ or 0.2 pc away from
distinct \ion{H}{2}
regions observed in radio~\citep{mor89,lan01b}. In particular, 
P114, P35, P100, P107 and
P112 are at the centers of the \ion{H}{2} regions named as H1,
H2, H12, H5 and H8, respectively, while  
P36 is 1\arcmin\ (2.4 pc) south of H1. The locations of
the stars, as well as the \ion{H}{2} regions and the Arches cluster,  
are labelled in Fig.~\ref{f:source_pos}. 

\subsection{Observations and Data Reduction}\label{ss:obs}
We used the 8-m \gem\ North telescope to obtain NIR spectra of the
stars (Program GN-2012A-Q-77). The $H$ and $K$ band spectra of 
P35, P98, P114 were observed with the spectrometer GNIRS. 
Its long slit (99\arcsec ) covered adjacent nebular emission, as well
as each of the three stars. But \gni\ was removed
from \gem\ for refurbishment in June, 2012, we thus switched 
to \nif\ and obtained only $K$ band
spectra of the other five stars. 

The \gni\ observations used the short camera with its 110.5 lines/mm grating and 
a slit width of 0.3\arcsec , resulting in a
spectral resolution of $\lambda/\delta\lambda$$\approx$5900 in the
$H$ and $K$ bands. The slit was centered at $\lambda$=1.7 $\mu$m ($H$
band) or 2.11 $\mu$m ($K$ band). The pixel scale along the slit was 0.15\arcsec
/pixel. We were able to select the roll angle, so that the slit covered not only 
the target stars, but also the peak emission of nearby \ion{H}{2} 
regions (see Fig.~\ref{f:source_pos}). The observing
pattern followed the standard `$ABBA$' dither sequence. The targets were
nodded along the slit ($A$ and $B$). 
The total exposure times are listed in Table~\ref{t:obs}. 
 Flat-field and wavelength calibration observations were
taken after each observation. Spectra of the standard
stars, HD 154663 (A1V) and HD
71254 (A1V),  were observed at air mass similar to the scientific
targets. The seeing of these observations was around 0.7\arcsec . 
According to the wavelength calibration files, the spectral resolution was
$\sim$2.9 \AA\ around 2.166 $\mu$m, corresponding to a velocity resolution of $\sim$40 km/s.

\nif\ is an Integral Field
Unit (IFU) with 3\arcsec$\times$3\arcsec\ field-of-view and has a 
spectral resolution of $\lambda/\delta\lambda$$\approx$5000. We used
the filter `HK\_G0603' and $K$ grating. The slit was centered at 2.2 $\mu$m. 
Nearby bright stars were used as natural guiding stars for the
AO. The
`$ABA$' dither pattern was used. For each target, we took two sets of
observations. Dark regions, as viewed in the \spitzer\ IRAC 3.6 $\mu$m
image~\citep{sto06}, 
were chosen as the off-source positions. The darkness of these
regions is due to the presence of foreground dense clouds. 
The exposure
time for each target is listed in Table~\ref{t:obs}. Flat-field images
were taken during the daytime and the wavelength calibration
observations were performed after each scientific observation. The
standard stars, HD 14606 (A0V) and HD 171296 (A0V), were observed at 
air mass similar to the scientific
targets. The spatial resolution of these observations was around 0.45\arcsec . 
According to the wavelength calibration files, we found that the spectral
resolution was 
$\sim$3.8 \AA\ around 2.166 $\mu$m, i.e. $\sim$53 km/s.

We followed the data reduction procedures listed in the examples
provided by the 
\gem\ \gni/\nif\ group, using the tasks contained in the \gem\
{\tt IRAF} \gni/\nif\ package. The
basic steps included trimming of the images, flat-fielding and sky
subtraction, as well as wavelength and distortion calibration. We defined the local sky
background for the sources at position $A$, as the emission
registered at position $B$ with the same detector coordinate. This
background was removed from the images at position $A$. The 
standard deviation in the wavelength calibration process was $\sim$0.1
pixel. This value was propagated into the uncertainty of the radial velocities determined
in \S\ref{ss:vel_s} and \S\ref{ss:vel_h}. Then, we extracted
the spectra of our scientific targets from the calibrated images, within
the central 0.75\arcsec . The same reduction procedures above were employed 
to extract
the spectra of the standard stars. We used the IDL package, {\tt
  xtellcor}~\citep{vac03},  
to remove the \ion{H}{1} absorption
lines from the spectra of standard stars to derive the telluric
absorption features. The corrected \gni\ and \nif\ spectra of the eight
stars are presented in Fig.~\ref{f:gnirs_spec} and
Fig.~\ref{f:nifs_spec}, respectively. In Fig.~\ref{f:nifs_spec}, we
also give the $K$ band spectra of the stars F8 (WN8-9h), F10 (O4-6 If$^+$) and
F15 (O4-6 If$^+$) in the Arches cluster~\citep{mar08}.

\section{Analysis and Results}\label{s:analysis}
\subsection{Spectral Classification of the Stars}\label{s:spec}
We compare
our new spectra with those presented in previous
works and classify our stars, mainly using the criteria from previous NIR spectroscopic studies of
massive stars by~\citet{mor96},~\citet{han05} and~\citet{mar08}.

\citet{cot99} present spectra (with $\lambda/\delta\lambda$=250) of six stars within or
near the \ion{H}{2} regions in the GC. These stars 
include P35 (H2:A), P107 (H5) and P112 (H8).  
Their low-resolution spectra clearly
show strong Br$\gamma$ 2.166 $\mu$m line. For P35 and P112, the
\ion{He}{1} 2.058 $\mu$m and \ion{He}{1} 2.112/3 $\mu$m
doublet are apparent. But because of the limited spectral 
quality,~\citet{cot99} only tentatively identify these stars as B[e]
stars.~\citet{mun06} update the spectral
type of P35 to O If$^+$,  based on their $H$-$K$ colors. 

\textbf{P107 and P35: }Our new high-resolution spectrum of P107 
shows additional diagnostic
features: a \ion{He}{1} 2.05 $\mu$m absorption
line, a \ion{He}{1}/\ion{N}{3} 2.112-2.115 $\mu$m emission complex
with a clear P-Cygni shape, a \ion{He}{2} 2.189 $\mu$m absorption
line and a very 
weak \ion{N}{3} 2.247/2.251 $\mu$m emission doublet. These features 
are typical characteristics of
both O If$^+$ stars and WN8-9h stars.~\citet{mar08} distinguish these
two spectral types mainly by the relative strengths of the Br$\gamma$
2.166 $\mu$m line and the 2.112-2.115 $\mu$m complex; for a WN8-9h star,
the Br$\gamma$ 2.166 $\mu$m line is stronger than the 2.112-2.115 $\mu$m 
complex, while for an O If$^+$ star, these two features have comparable
strengths. Based on this criterion, we would categorize P107 as 
a WN8-9h star. However, WN8-9 stars should 
also show a signifcant 2.112-2.115 $\mu$m complex and 
\ion{He}{2} 2.189 $\mu$m
absorption lines with a P-Cygni shape (see the spectra of F8
 in Fig.~\ref{f:nifs_spec}). The later feature is abscent in
the spectrum of P107. On the other hand, like star 9 and 10 
in~\citet{mau10c}, the P-Cygni shape of the 2.112-2.115 
$\mu$m complex of P107 is similar to the O
If$^+$ stars, F10 and F15, in the Arches cluster (see
Fig.~\ref{f:nifs_spec}). 
Thus, we classify P107 as an O4-6
If$^+$ star, instead of WN8-9h.   
The \ion{He}{2} 2.189 $\mu$m absorption line without 
P-Cygni shape but with a weak 2.112-2.115
 $\mu$m complex emission is also found in P35, 
consistent with its classification as an O If$^+$ star by~\citet{mun06}. 

\textbf{P112: }We still keep the B[e] assignment for P112 
by~\citet{cot99} , although
we cannot exclude the possibility that it is a LBV star. Its \nif\ 
$K$ band spectrum is very similar to that of the 
LBV star, `S Dor'~\citep{mor96}. We find 
\ion{Fe}{2} 2.089 $\mu$m and \ion{Mg}{2} 2.138/2.144 $\mu$m lines, 
but no CO
overtone (2.3-2.4 $\mu$m), which is typically found in B[e] spectra due to
the presence of stellar disks~\citep{mor96}. 
However, P112 and GCIRS34W, another LBV candidate~\citep{tri06}, have similar
$K$ band magnitudes, but they are approximately three magnitudes
fainter than the three known LBVs in the GC, Pistol, qF362 and 
G0.120-0.048 ($m_K<$7.5
mag,~\citealt{mau10b}). This difference is much greater than 
what may be expected from
typical variability of LBV stars 
(which is $\sim$1 magnitude, e.g.~\citealt{mau10c}). 
Unfortunately, P112 is below the
detection limit of~\citet{gla01} ($m_K<$10.5 mag in the survey field
`GC2') and we do not have available information about its
variability. Future monitoring program is needed
to further constrain its spectral type. 

\textbf{P36 and P114: }they are first considered as evolved massive
star candidates, because of their detection (just like P35) in both Chandra and 2MASS
observations~\citep{mau09}. \textbf{Follow-up spectroscopic observations further classify them 
into O4-6 I stars~\citep{mau10a,mau10c}. Our new high sensitivity \nif\ spectrum of 
P36 shows that there is 5-7\% enhancement near the Br$\gamma$ 2.166 
$\mu$m. Therefore, we reclassify it as an O If$^+$ star~\citep{han05}, as suggested by the 
referee.} The Br$\gamma$ 2.166 $\mu$m absorption
line appears stronger and narrower in our spectrum of P114 
than in that of Mauerhan et al. (2010a), which could be
the result of our over-subtraction of local background, i.e., 
contamination by hydrogen emission from the nearby \ion{H}{2} 
region (see \S~\ref{sss:vel_g}). Therefore,
we exclude the hydrogen lines in the measurement of the radial velocity of P114 in
\S~\ref{ss:vel_s}. 

\textbf{P97, P98 and P100: }\citet{don11} identify these three sources as
PESs. The Paschen-$\alpha$
equivalent widths and intrinsic intensities at
1.90 $\mu$m of the sources indicate that they are OB supergiants (see Fig. 13 of~\citealt{don12}). \citet{mau10c}
spectroscopically confirm that P98 and P100 are O4-6 If$^+$ stars. However, our new GNIRS spectrum 
does not have \ion{He}{2} 2.189 $\mu$m absorption line, but strong \ion{He}{1} 1.700 $\mu$m absorption 
line. Therefore, P98 could be a later type, such as O9-B0 If$^+$ star~\citep{han05}. Our 
\nif\ spectrum of P97 is similar to that of
P107 and F15 in~\citet{mar08}. We thus 
classify P97 as an O4-6 If$^+$ star.  

\subsection{Color Magnitude Diagram}\label{s:age}
We show the extinction-corrected color magnitude diagram in Fig.~\ref{f:iso}. 
The NIR magnitudes for the majority of
our stars are from the SIRIUS catalog\footnote{Simultaneous three-colour
InfraRed Imager for Unbiased Surveys (SIRIUS) is mounted on the 
Infrared Survey Facility (IRSF) in South Africa~\citep{nag03}. This
camera surveyed the region $|l|$ $<$ $2^{o}$ and $|b|$ $<$ $1^{o}$,
with an angular resolution of $\sim$1.2\arcsec\ in the $J$ 
band.}~\citep{don12}. The SIRIUS catalog did not report magnitudes for
 P97 and P112; so for these sources we use the information from the recent 
UKIRT GC Plane survey~\citep{luc08}\footnote{Although the throughputs of the $J$ and $H$
  bands in the UKIRT and SIRIUS detectors are similar, the effective
  wavelength of the $K$ band (2.20 $\mu$m) in the UKIRT survey is longer
  than that of the SIRIUS survey (2.14 $\mu$m).}. 
We convert the UKIRT
magnitudes into the SIRIUS magnitudes using Eqn. 3-5 in~\citet{hew06}
and Eqn. 7-9 in~\citet{don12}. The foreground
extinctions are from Table 2 of~\citet{don12}, which
are derived from the NIR colors of individual stars. We adopt the 
extinction curve of~\citet{nis06} to convert 
$A_{F190N}$ to $A_K$ ($A_K$=0.787$A_{F190N}$, see
Table~\ref{t:sample}) and to obtain $A_J=3.021A_K$. The
high extinctions ($A_K$$>$1.95) of our program stars are consistent
with their being in the
GC. 
Fig.~\ref{f:iso} also includes the OB supergiants in the Arches cluster for comparison. 

In Fig.~\ref{f:iso}, we also overlay the isochrones 
of four stellar ages as representative stellar
populations and the evolutionary tracks of stars
with four masses from the Geneva
model~\citep{eks12}\footnote{http://obswww.unige.ch/Recherche/evoldb/index/Isochrone/}. We
assume solar metallicity and no rotation. In Fig.11 of~\citet{don12}, the $H$-$K$ colors of the 
OB supergiants and O If$^+$ stars are bluer than the WN stars, but are similar to those of the field stars from \spitzer/IRAC 
GALCEN survey~\citep{sto06}, which mainly are Red Clump and Red Giant Branch stars. Therefore, the contribution 
of free-free emission from the wind, which is significant in the $K$-band of WR stars~\citep{mau10a} 
should be negligible 
in the OB supergiants and O If$^+$ stars. Therefore, no free-free emission is considered here.

We estimate the ages and masses of our
program stars based on their locations in the color-magnitude diagram. 
For each star, we use the following equation to estimate 
the maximum likelihood of a specific age and mass based on 
a comparison of the observed and model colors and 
magnitudes~\citep{bai11}: 
\begin{equation}
\rho (Age,Mass) = exp\langle-\frac{[JK-JK_m(Age,Mass)]^2}{2\delta JK^2}-\frac{[K-K_m(Age,Mass)]^2}{2\delta K^2}\rangle
\end{equation}
where $JK$ and $K$ are the extinction-corrected $J$-$K$ colors and $K$-band
magnitudes. The subscript `m' indicates the values predicted
from the Geneva model as a function of stellar age and mass.
 $\delta JK$ and $\delta K$ are the corresponding uncertainties in  
$JK$ and $K$. The estimated age and mass, as well as their errors, are listed in
Table~\ref{t:sample}. 

\subsection{Radial Velocity Measurements of Stars}\label{ss:vel_s}

Table~\ref{t:velocity} lists the spectral features used to measure the radial
velocities of individual stars. We mainly use the absorption lines, 
\ion{He}{2} 10-7 2.189 $\mu$m in the $K$ band and \ion{He}{2} 12-7 1.693 $\mu$m in the 
$H$ band in particular, to 
determine the radial velocities of the stars. But, we also use 
the \ion{N}{3} 2S-2Po 2.247/2.251 $\mu$m emission doublet, if present\footnote{Compared to 
the \ion{He}{2} lines, the \ion{N}{3} doublet are less affected by the stellar winds. 
We derive the radial velocities from \ion{He}{2} 10-7 2.189 $\mu$m 
and \ion{N}{3} 2S-2Po 2.247/2.251 doublet, respectively, for P107,
 P100, P97 and P36, and find that they are different by less than 20 km/s and 
 are consistent within the uncertainty. Because P107 and P100 have the                                                                                                                                                                                                                                                                          
 strongest Br$\gamma$ lines in our O supergiant and O If$^+$ stars, i.e. the strongest stellar 
 wind, we believe that in our sample, like \ion{N}{3} doublet, the \ion{He}{2} lines are less affected by the wind too.}.
These absorption lines and emission doublet are relatively isolated
and arise from regions very 
close to the stellar photospheres due to non-LTE effects\footnote{The 
\ion{C}{4} 2.08 $\mu$m triplet also supposedly arise from the
  photospheres. However, we find that the radial velocities derived
  from this triplet are 
  blueshifted by $>$100 km/s, compared with those from \ion{He}{2} 10-7
2.189 $\mu$m and the \ion{N}{3} 2S-2Po 2.247/2.251 $\mu$m
doublet, assuming the vacuum wavelengths in the Atomic Line list (v2.05). We suspect
that the wavelengths may be problematic (Francisco Najarro, private
communication) and thus do not
include the triplet in the radial velocity measurement.}   
(Fabrice Martins, private communication). If these distinct spectral 
features are totally absent, we then use another atomic
hydrogen line, \ion{He}{1} $4d~^3D-3p~^3P^o$ 1.70 $\mu$m, or metal lines 
instead (see Table~\ref{t:velocity}). We give the vacuum wavelengths
of these lines in 
Table~\ref{t:vacuum}. In Appendix A, we further 
discuss the effects of stellar binarity on the velocity measurements.




We use the combination of a Gaussian function and a linear equation to
fit the absorption or emission line and the adjacent normalized
continuum through the least $\chi^2$ fitting method. 
When multiple lines are presented in the same spectrum,
they are jointly fitted with their centroid wavelength ratios fixed to
the NIR rest-frame values in the
vacuum and the same velocity dispersion, which are 
linearly related to the standard deviation of the Gaussian function.  
The exception is P112, because the Br$\gamma$ 2.166 $\mu$m 
and \ion{Fe}{2} 2.089 $\mu$m originate in different parts of
the expanding stellar winds. The radial velocities derived from these two
lines should be considered as the lower limit
for P112. In Fig.~\ref{f:p97_fit}, we show the fitting results for the \ion{He}{2} 10-7
2.189 $\mu$m absorption line and the \ion{N}{3} 2S-2Po 2.247/2.251 $\mu$m emission 
doublet of P97, which has the strongest \ion{N}{3} doublet in our sample, 
to demonstrate our fitting result. The Guassian function fit the lines very well and 
we do not see any asymmetry in the line profile of \ion{He}{2} 10-7
2.189 $\mu$m, which could be potentially caused by the stellar wind.
We use the routine `{\tt MPFIT}'~\citep{mar09} to
find the best-fit parameters and use the Monte Carlo method to derive
the uncertainty. This involves randomly adding a value that follows a normal
distribution with a mean of zero and a standard deviation equal to
the intensity uncertainty, to the observed intensity at each spectral pixel. Then we 
fit the new spectrum and obtain a new velocity and dispersion. Repeating this 
process 10$^4$ times,  the 68\% percentiles of the
outputs are the errors of our parameters. In
Table~\ref{t:velocity}, we list the velocities and dispersions for the
eight sources.

\subsection{Radial Velocity Measurements of \ion{H}{2} regions}\label{ss:vel_h}
\subsubsection{GNIRS}\label{sss:vel_g}
Figs.~\ref{f:p35_spec}-\ref{f:p114_spec} show the locations of the slit
on the \hst/NICMOS F190N and 
Paschen-$\alpha$ images for P35, P98 and
P114. The figures also include the velocity-position diagrams of the
Br$\gamma$ 2.166 $\mu$m line for these three stars and the ionized surface of nearby molecular clouds.

In Figs.~\ref{f:p35_spec}c-\ref{f:p114_spec}c, we present the
  `continuum-subtracted' spectra of these three stars. We use 
the `line-free' spectra at 2.155-2.158 $\mu$m and
  2.174-2.177 $\mu$m to interpolate the continuum at
  2.158-2.174 $\mu$m, which is subtracted from our
  \gem/\gni\ spectra for the line images around 
  Br$\gamma$ 2.166 $\mu$m.  Consequently, most of the stellar continuum
  emission in the field-of-view has been
eliminated. We also convert the units of the ordinate from
  wavelength ($\mu$m) to radial velocity (km/s), relative to the
  rest wavelength of 
  Br$\gamma$ 2.166 $\mu$m. 
We extract the spectra from
the regions within the thick red lines in Figs.~\ref{f:p35_spec}c-\ref{f:p114_spec}c to measure the
radial velocities of the nearby \ion{H}{2} regions. We use
the same method in \S~\ref{ss:vel_s} to derive the centroids and their
uncertainties for the Br$\gamma$ 2.166 $\mu$m emission, from which we derive the radial velocities, given in Table~\ref{t:hii} and Figs.~\ref{f:p35_spec}d-\ref{f:p114_spec}d. 
 

\subsubsection{NIFS}\label{sss:vel_n}
Due to the small field-of-view of \nif , we can only study the
emission from the adjacent \ion{H}{2} region within 
3\arcsec$\times$3\arcsec\ of each source. For the five stars with
\nif\ observations, only P112 has 
extended hydrogen emission within the \nif\ field-of-view, 
as seen in our \hst/\nicmos\ Paschen-$\alpha$ image. 

In Fig.~\ref{f:p112}, we show the \hst/NICMOS F190N continuum image
and 
Paschen-$\alpha$ emission image of P112 at the same spatial scale as
the \nif\ observations; the \spitzer\ \irac\ 3.6 $\mu$m~\citep{sto06} 
image of the region; and \nif\
Br$\gamma$ images for three velocity ranges ([-85,-25], [-25,35] and [35,
95] km/s) for P112. 
In the \hst\ Paschen-$\alpha$ image, we see an enhancement in
the $\sim$1.1\arcsec\ (0.045 pc) to the northwest of P112, as well as a
faint ring structure to the southwest. This emission is not attributable to stellar
emission, as can be seen by comparison to the \hst\ F190N image.
 In the \spitzer\ \irac\
3.6 $\mu$m image, we notice that there is an infrared dark cloud
(IRDC) to the west of P112. Therefore, the Paschen-$\alpha$ 
emission could come from the
surface of this cloud, suggesting that P112 and the IRDC are physically
related. For our \nif\ data, we subtract the stellar spectra
around Br$\gamma$ 2.166 $\mu$m using continuum measurements 
inferred from the surrounding wavelength ranges 2.155-2.158 $\mu$m and
  2.174-2.177 $\mu$m. In the bottom panel of
  Fig.~\ref{f:p112}, we show the \nif\ images at three velocity
  ranges. Weak emission from the northwest of P112 is evident 
  within the [-25,35] km/s range. We extract the spectra from this region to
  more accurately derive the radial velocity, including the
  uncertainty, for the ionized gas. This velocity is presented in Table~\ref{t:hii}.

\section{Discussion}\label{s:dis}
The above results enable us to explore the origins of 
our program stars. In
\S\ref{ss:comparison}, we examine  
the relationship of our program stars to their adjacent ionized/molecular
gases, through a comparison of their radial velocities. In \S\ref{ss:int}, we
provide an in-depth discussion about the link between the PESs
P114/P35 and the shell-like \ion{H}{2} regions, H1/H2, respectively. 
In \S\ref{ss:ori}, we
examine the possible mode of star formation in the GC.  

\subsection{Radial Velocity Comparison}\label{ss:comparison}
Fig.~\ref{f:los_com} compares the velocity measurements of our program 
stars and their adjacent diffuse gases as
a function of their projected distance from the Arches cluster. The figure 
includes independent velocity measurements of ionized and 
molecular gases from previous radio observations of the H92$\alpha$ 
recombination
line~\citep{lan01b,lan02} and molecular lines (e.g., CS,
HCN, ~\citealt{tsu99,tsu11}). The shaded area represents
the one sigma uncertainty range of the radial velocity of the Arches
cluster (87$-$103 km/s)~\citep{fig02}. The radial velocity of the
Quintuplet cluster (102$\pm$2 km/s)~\citep{lie09,sto14} is very
similar to that of the Arches cluster and has not been marked in
Fig.~\ref{f:los_com}. Clearly, the Arches and Quintuplet
clusters are redshifted by at least 50 km/s relative
to all eight of our program stars. 

The radial velocities of P35 and P114 
are similar to that of the nearby -30$-$0 km/s
 molecular cloud~\citep{tsu99}. Fig.~\ref{f:p35_spec}d
 and~\ref{f:p114_spec}d show some velocity gradients of the ionized gas
 near the stars at $>$3 sigma level. At projected distances of 
$\sim$ -5\arcsec\ (0.2 pc) and $\sim$30\arcsec\ (1.2 pc) from 
the stars, the ionized gas is blueshifted, as compared to both the
stellar velocities and that of the nearby molecular gas. However, 
the gradients between
the vertex of the \ion{H}{2} regions and P35/P114 stars are
different. We show zoomed-in versions of Fig.~\ref{f:p35_spec}c
 and~\ref{f:p114_spec}c in Fig.~\ref{f:p35_p114}. Moving toward the program stars, the radial
velocity of the ionized gas decreases from about -80 km/s to
0 km/s in H2, while it increases from -40 km/s to -70 km/s in 
H1. Because P114 and
P35 are the only bright evolved massive stars
near the H1 and H2 \ion{H}{2} regions, these observed strong velocity
gradients support the argument presented in~\citet{don12} that
P114/P35 are physically associated with their \ion{H}{2} 
regions. We explore the relationship
between P114/P35 and H1/H2 further in
\S\ref{ss:int} and \S\ref{ss:ori}. 

Unlike P35 and P114, the 
radial velocities of P107, P97 and P112 are significantly different from those of
ionized/molecular gases that are nearby in projection. P107 is located
within the gap of the two parts of the H5 \ion{H}{2} region 
(Fig.~\ref{f:source_pos}; see also Fig. 15 of~\citealt{don12}) and is
redshifted relative to the nearby
ionized gas (-34.4$\pm$35.8 km/s;~\citealt{lan01b}) and molecular
clouds ($\sim$-30$\pm$20 km/s;~\citealt{ser87}). Like P107, the radial
velocities of P97 and the apparently nearby G0.01+0.02 \ion{H}{2}
region (ionized gas: -28$\pm$42.4 km/s, molecular gas: $\sim$-20$\pm$10 km/s;~\citealt{lan01b}) 
differ by more than 60 km/s. 
The lower limit of the radial 
velocity of P112 (40.1$\pm$5 km/s) is significantly 
larger than that of the ionized
gas (6.5$\pm$3.4 km/s). Based on 
location and derived stellar types these stars could be the 
ionizing sources of the projected nearby \ion{H}{2} regions. Therefore,
the stars are more 
likely just passing near the molecular clouds, 
than emerging from them. 

P98 and P100 lie near the arched radio filaments. The radial velocities of
the nearby ionized gas derived from the Br$\gamma$ 2.166 $\mu$m
measured from our
spectra and the H92$\alpha$ values of \citet{lan01b} (-43.6$\pm$22.6 
km/s) are
similar to that of P98.  If the ionized gas near P100 is an extension 
of the Arched
Filaments, we would expect its radial velocitiey to be 
$\sim$-35 km/s (e.g. Fig. 7 
of~\citealt{lan01b}, where P100 would be located near boxes
 `15', `16'
and `17'). Assuming this velocity for the gas, it is slightly
blueshifted, when compared to P100. However, unlike P35 and P114, 
in our \hst/NICMOS Paschen-$\alpha$ image, 
P98 is not embedded within the nearby \ion{H}{2} region, 
G0.07+0.04, which is attributable to the Arches cluster, and we do not find a radial-velocity gradient 
in the \gni\ spectra of the ionized gas near the star. In addition, 
the morphologies of ionized gas near 
P98 and P100 do not indicate that they
formed {\sl in-situ}. Therefore, P98 and P100
are probably not 
physically associated with the projected nearby \ion{H}{2} regions. 

Unlike the other seven stars, P36 is relatively isolated, with no
obvious nearby \ion{H}{2} region. The measured radial velocity for
this star is slightly redshifted, when 
compared to the nearby -30$-$0 km/s 
molecular cloud. The data suggest that P36 formed elsewhere and
recently approach  
the -30$-$0 km/s 
molecular cloud in projection.

\subsection{The Cometary Morphology of the H1 and H2 \ion{H}{2} regions}\label{ss:int}
\begin{figure*}[!thb]
 \centerline{
      \epsfig{figure=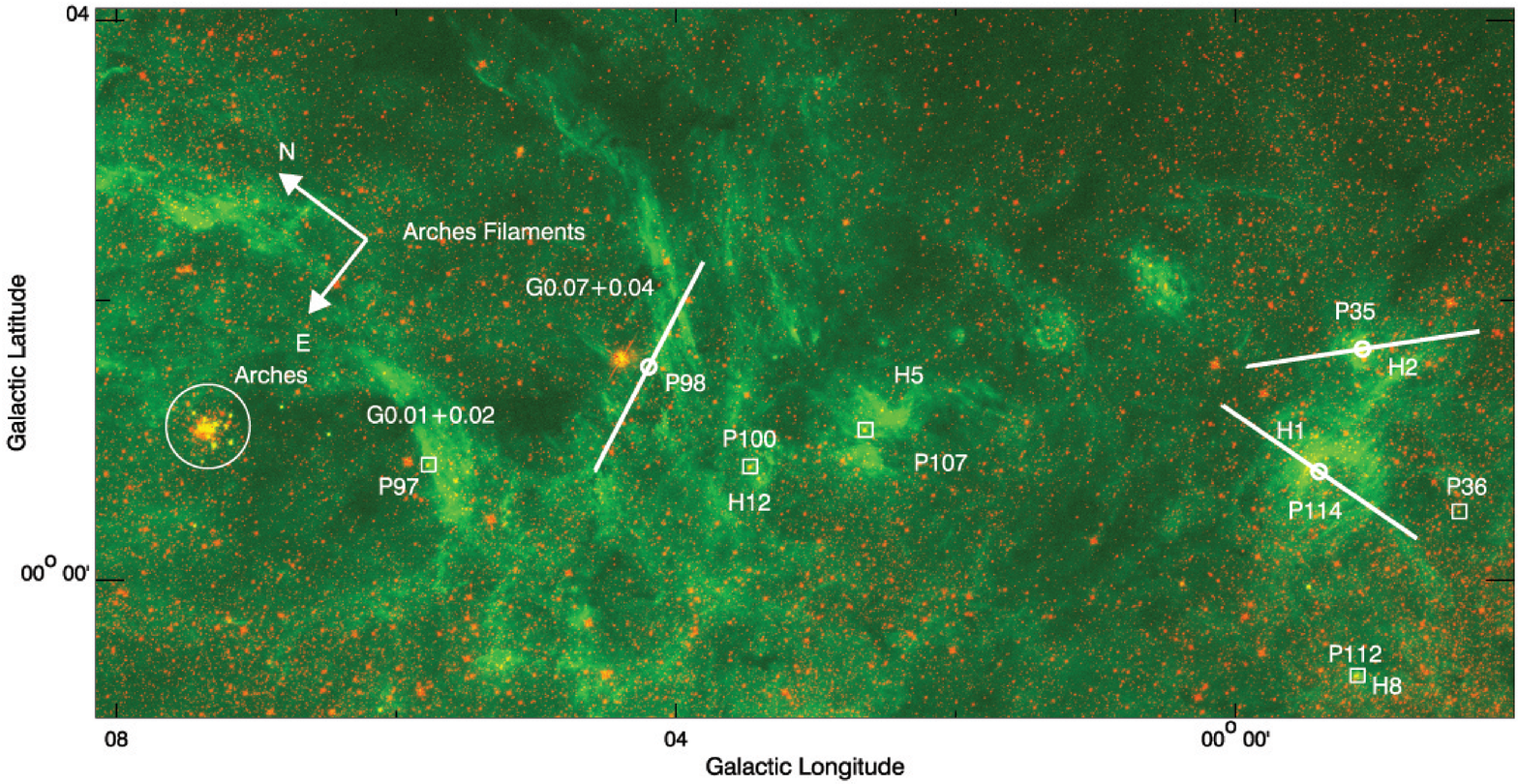,width=1.0\textwidth,angle=0}
}
 \caption{Eight evolved massive stars studied in this paper overlaid on
   the color image, where red traces the stellar continuum intensity
   in the \hst\ F190N filter, while green represents Paschen-$\alpha$ 
 emission at 1.87 $\mu$m~\citep{wan10,don11}. The
 locations of the Arches cluster and \ion{H}{2} 
 regions are also labelled. The `circles' (boxes) mark the
 sources observed with \gem\ \gni\ (\nif ), while the white lines
 represent the coverages of the \gem\ \gni\ long slit.}
 \label{f:source_pos}
 \end{figure*}

\begin{figure*}[!thb]
  \centerline{
      \epsfig{figure=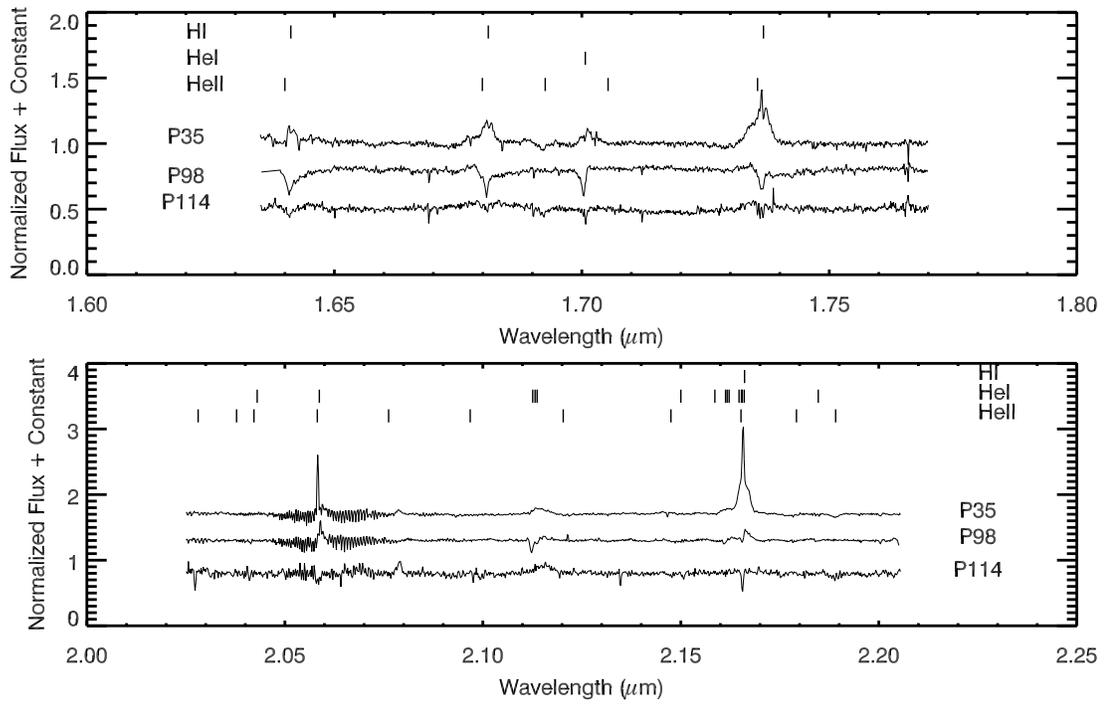,width=1\textwidth,angle=0}
}
 \caption{\gem\ \gni\ spectra of P35 (O If$^+$) , P98(O9-B0 If$^+$) and P114
   (O4-6 I) in $H$ band (top
   panel) and $K$ band (bottom panel).} 
\label{f:gnirs_spec}
 \end{figure*}

\begin{figure*}[!thb]
  \centerline{
      \epsfig{figure=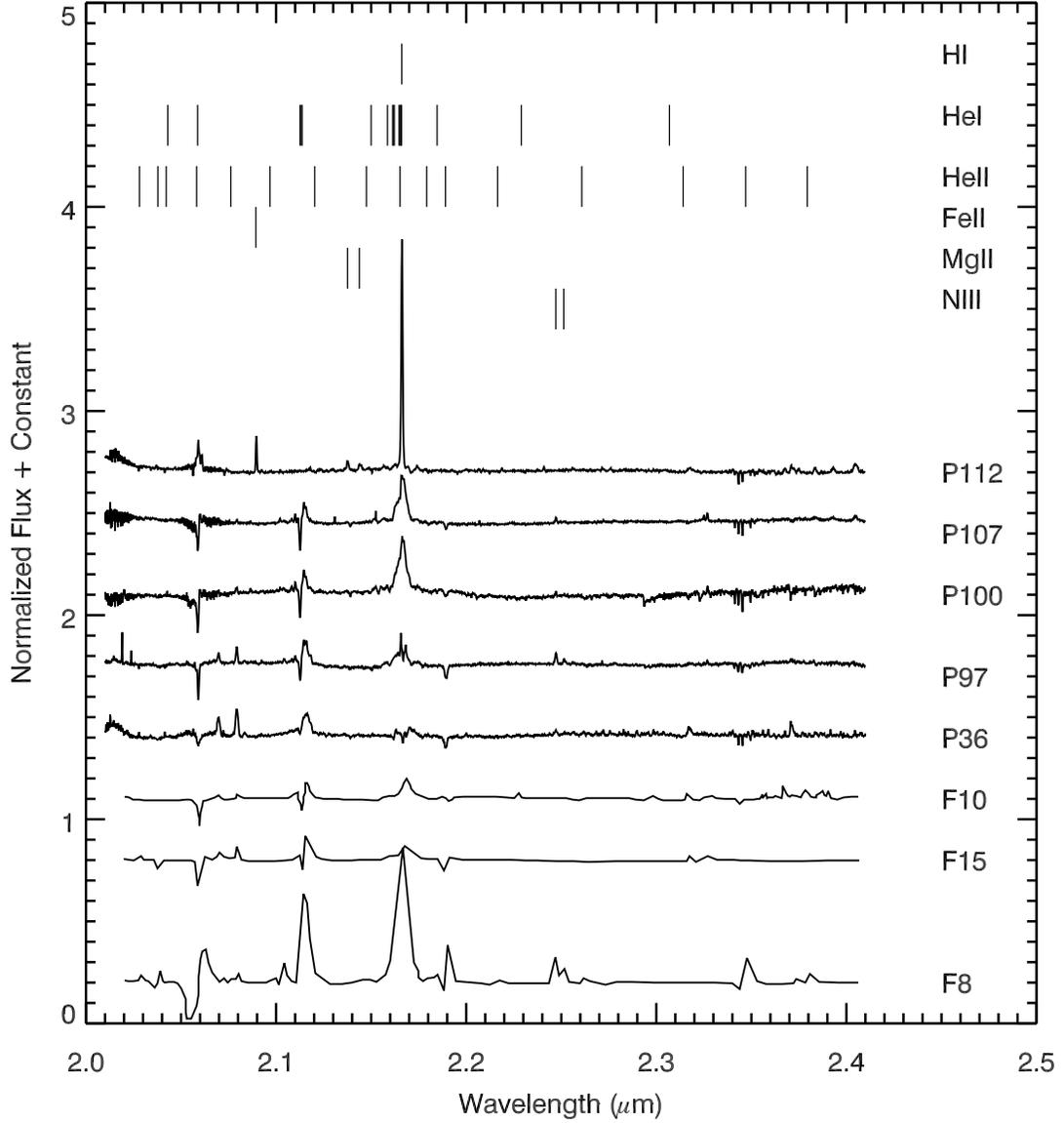,width=1\textwidth,angle=0}
}
 \caption{\gem\ \nif\ $K$ band spectra of P36 (O If$^+$),
   P97 (O If$^+$), P100 (O4-6 If$^+$), P107 (O4-6 If$^+$) and P112
   (B[e]), as well as the $K$ band spectra of F8 (WN8-9h), F10 (O4-6 If$^+$) and
 F15 (O4-6 If$^+$), members of the Arches cluster~\citep{mar08}.} 
\label{f:nifs_spec}
 \end{figure*}

\begin{figure*}[!thb]
  \centerline{
      \epsfig{figure=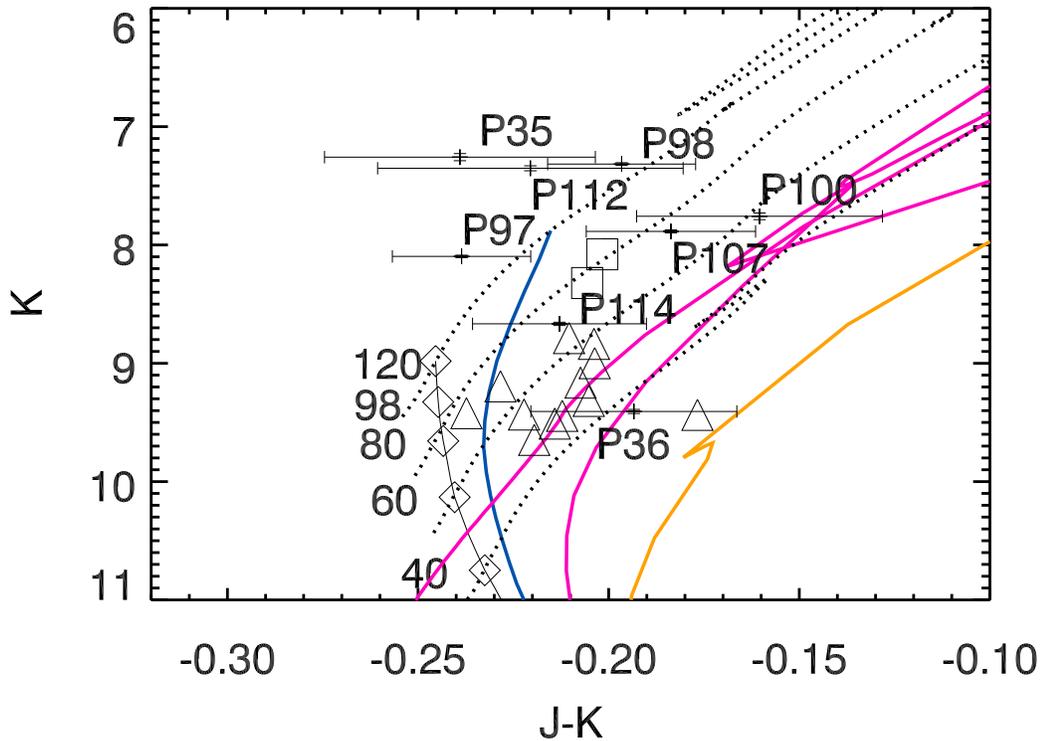,width=1\textwidth,angle=0}
}
 \caption{Extinction-corrected color magnitude diagram, $J$-$K$ vs
   $K$. The data points with error bars represent the measurements for our eight sources. 
   The black, blue,
   pink and yellow solid lines are the Geneva isochrones (assuming solar
   metallicity) at ages of 1, 2, 4 and 6.3 Myr, respectively. The
   diamonds mark the location of stars with 40, 60, 80, 98 and 120
   M$_{\odot}$ at 1 Myr isochrone. The dotted
   lines are the evolutionary tracks of stars with 40, 60, 85 and 120
   M$_{\odot}$. The
 `boxes' and `triangles' represent the known O If$^+$ and O supergiants in the
 Arches cluster.}
\label{f:iso}
 \end{figure*}

\begin{figure*}[!thb]
  \centerline{
      \epsfig{figure=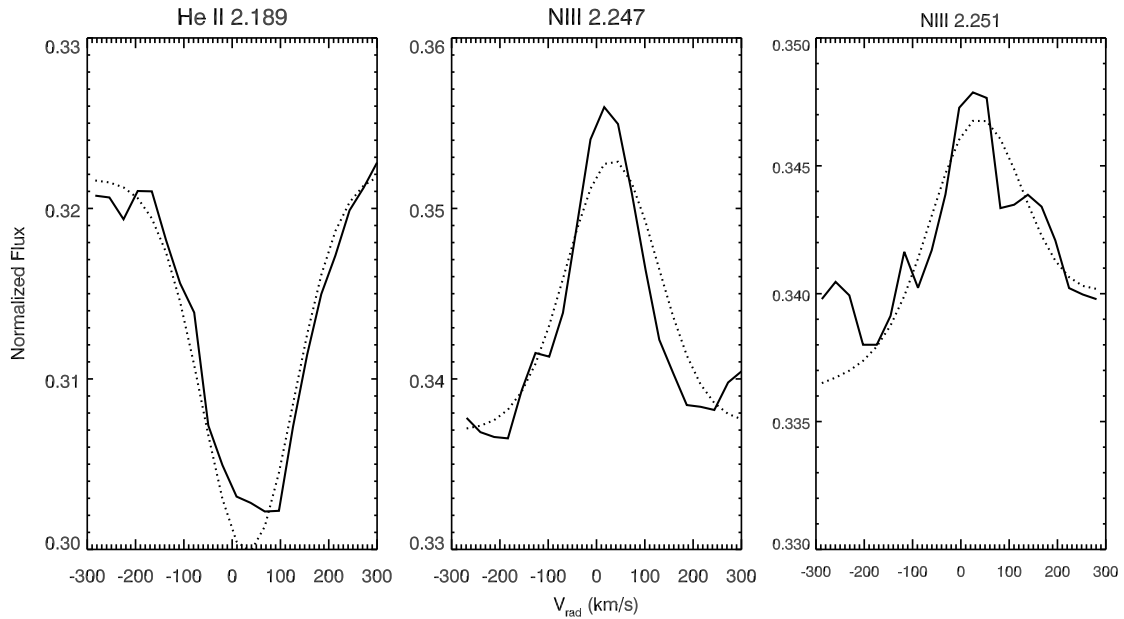,width=1\textwidth,angle=0}
}
 \caption{Fitting the \ion{He}{2} 2.189 $\mu$m absorption line 
 and \ion{N}{3} 2.247/2.251 $\mu$m emission doublet of P97. The solid lines 
 are the observed spectra, while the dashed lines are the Gaussian 
 function, plus a linear 
 equation to represent the underlying stellar continuum. The abscissa 
 is the radial velocity, with respect to the vacuum 
 wavelengths of individual lines. The \ion{He}{2} 2.189 $\mu$m absorption 
 line (43.8$\pm$17.9 km/s) is slightly redshifted relative to 
 \ion{N}{3} 2.247/2.251 $\mu$m emission doublet (26.2$\pm$15.2 km/s), but within the uncertainty range. 
 In \S\ref{ss:vel_s}, we fit these two lines simultaneously, in order 
 to reduce the uncertainty of the radial velocity.} 
\label{f:p97_fit}
 \end{figure*}

\begin{figure*}[!thb]
 \centerline{
      \epsfig{figure=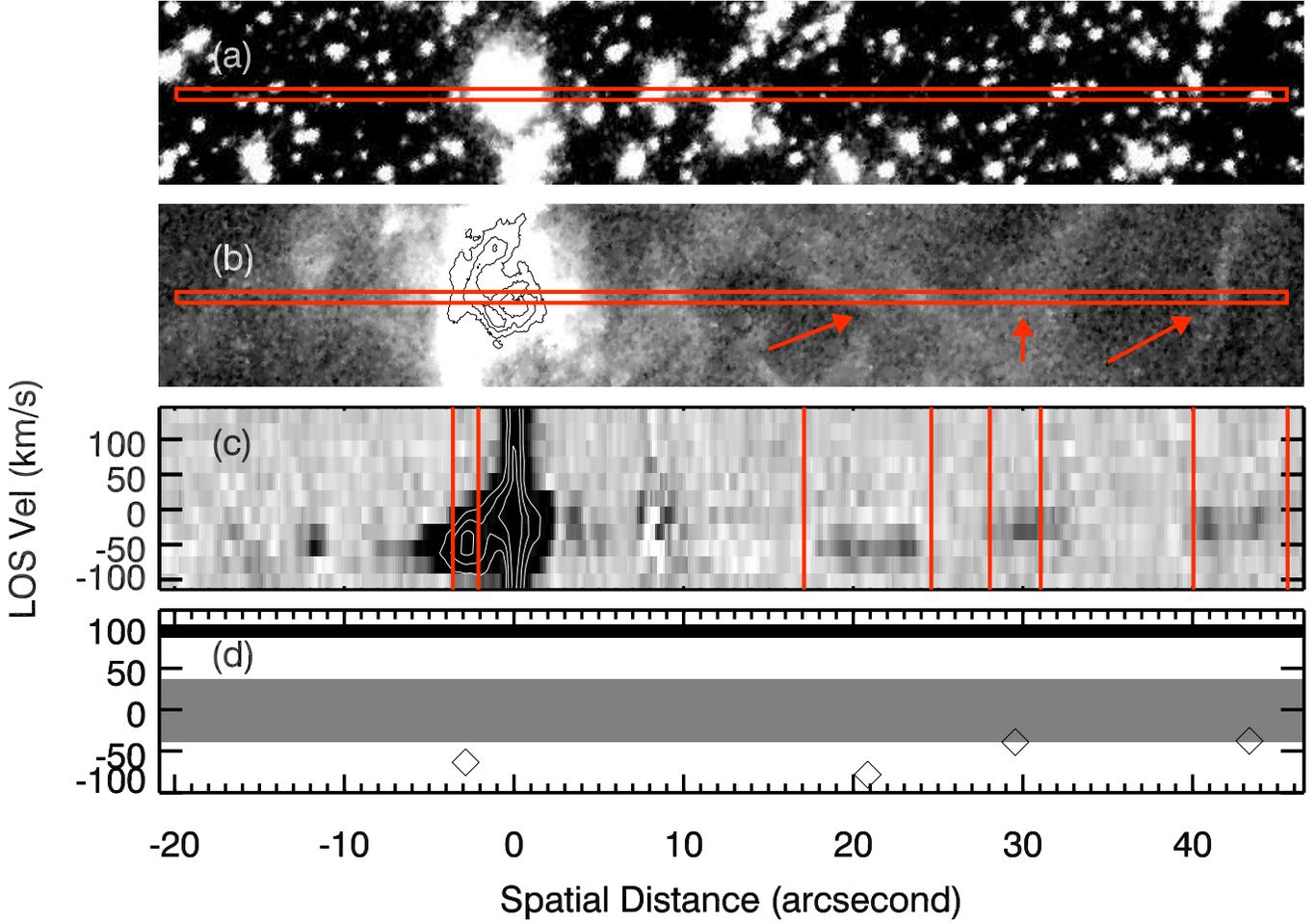,width=1.2\textwidth,angle=0}
}
 \caption{NIR imaging and spectroscopic views of P35 and H2. (a) and
   (b) are the F190N and Paschen-$\alpha$ images from~\citet{don11}. 
   The long slit used to obtain the spectrum is outlined in panels (a)
   and (b). The red arrows in the Paschen-$\alpha$ image mark the
   low-surface brightness emission features clearly seen in the
   spectrum. (c) the velocity-position plot obtained from the spectrum
   of the Br$\gamma$ line. We extract the spectra from the regions 
bracketed by red thick lines to determine the radial velocities of
low-surface brightness regions. (d) comparison of the radial velocities of P35 and the
   features with that of the Arches cluster (98$\pm$8 km/s; the
   shaded area). The abscissa of these plots is the projected distance
   relative to P35 along the long slit in units of
   arcseconds. The diamonds represent the radial velocities of ionized
   gas at different locations along the long slit, the uncertainty of
   which are much smaller than the symbols.}
\label{f:p35_spec}
 \end{figure*}

\begin{figure*}[!thb]
  \centerline{
      \epsfig{figure=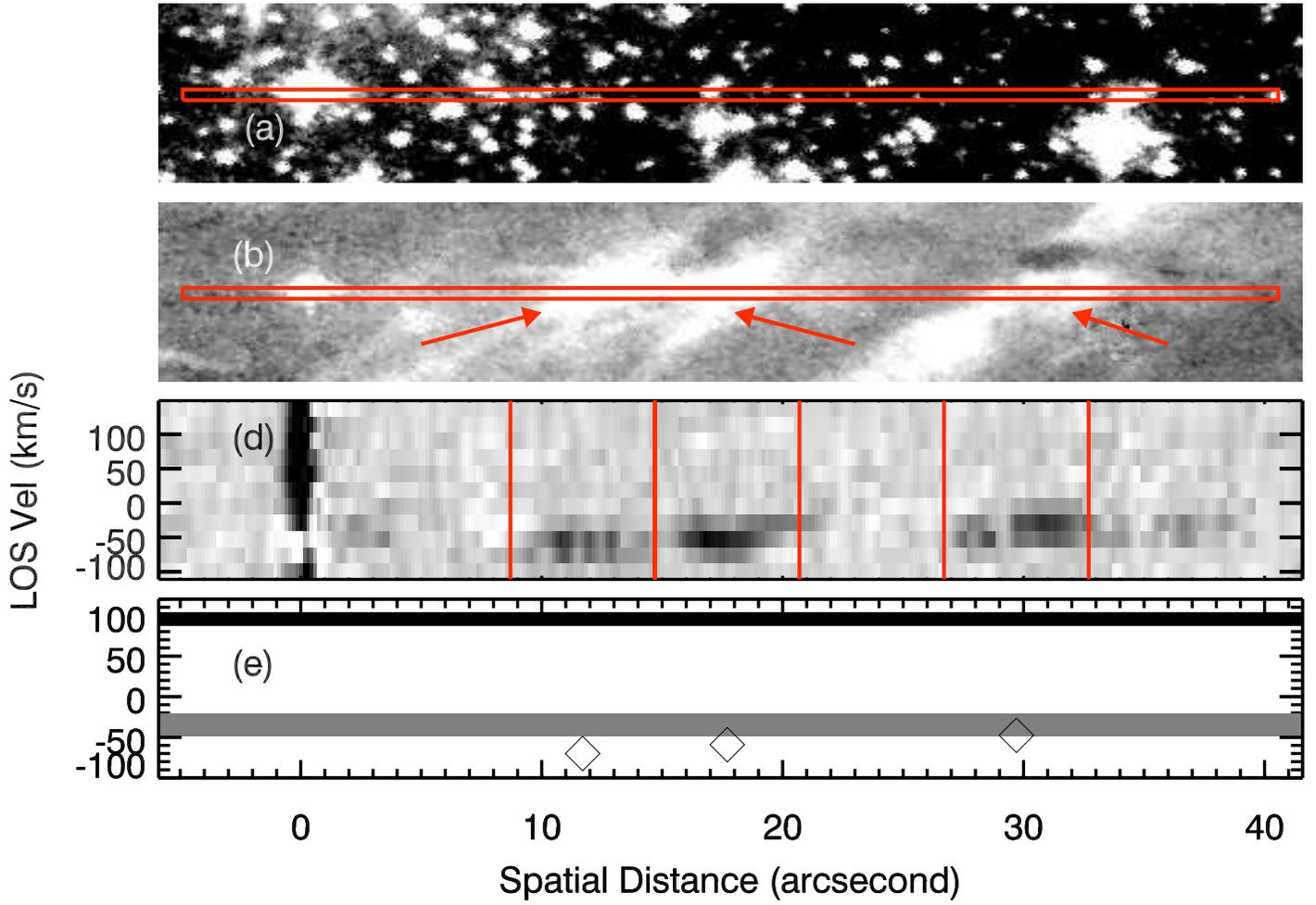,width=1.2\textwidth,angle=0}
}
 \caption{Same as Figure~\ref{f:p35_spec}, but for P98.}
\label{f:p98_spec}
 \end{figure*}

\begin{figure*}[!thb]
  \centerline{
     \epsfig{figure=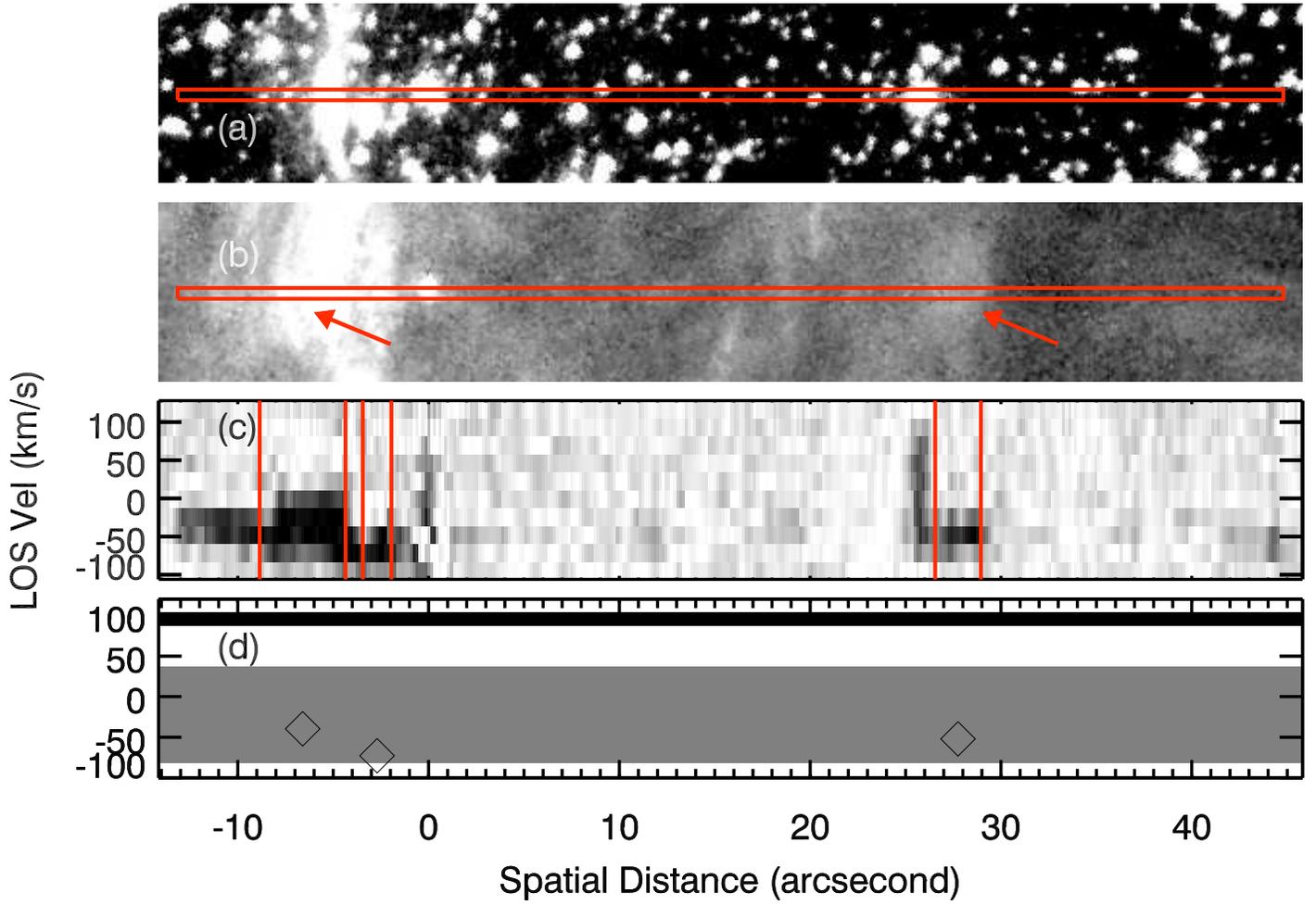,width=1.2\textwidth,angle=0}
}
 \caption{Same as Figure~\ref{f:p35_spec}, but for P114.}
\label{f:p114_spec}
 \end{figure*}

\begin{figure*}[!thb]
  \centerline{
      \epsfig{figure=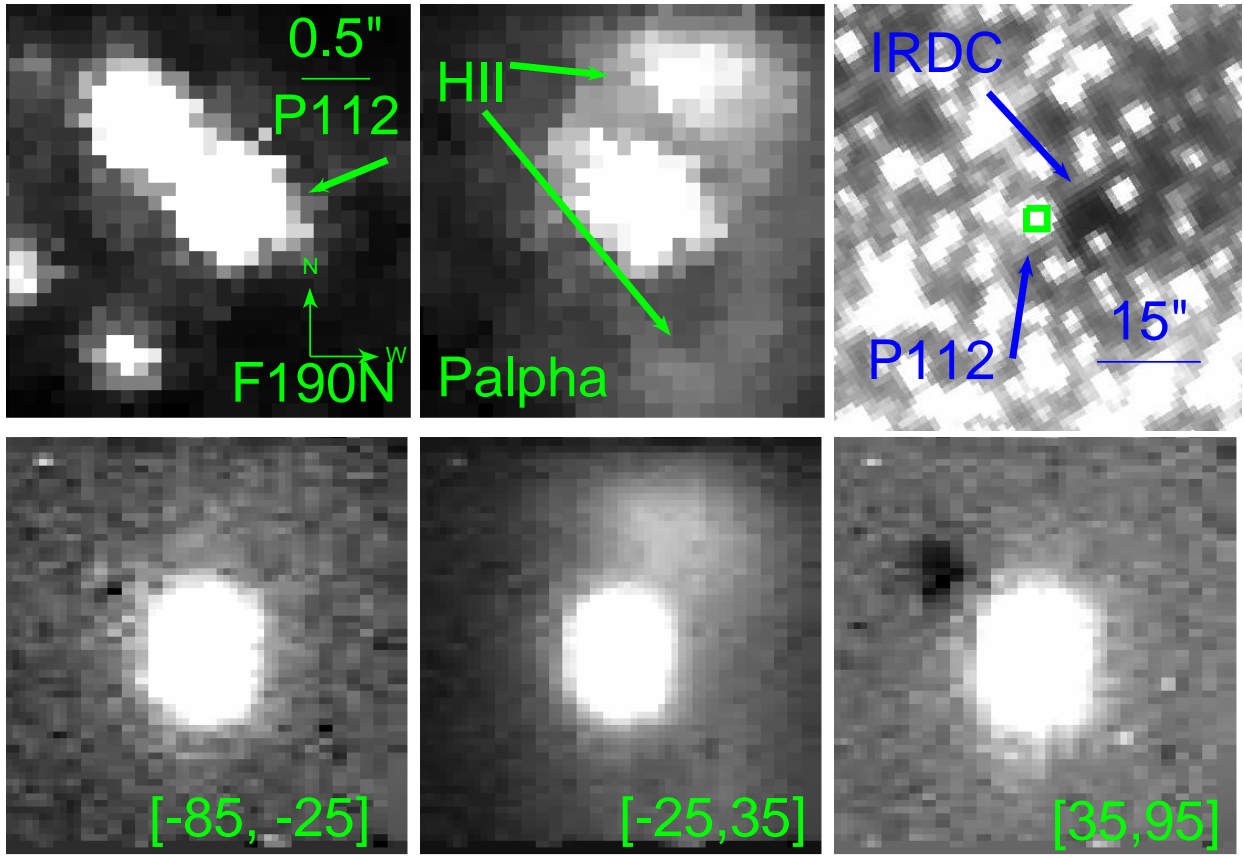,width=1\textwidth,angle=0}
}
 \caption{Images of P112 and its immediate vicinity. 
Upper panel, from left to right: F190N and 
   Paschen-$\alpha$ images from the HST/NICMOS survey~\citep{wan10}
   and \spitzer\ \irac\ 
   3.6 $\mu$m image from~\citet{sto06}. The green box
   in the \spitzer/\irac\ image outlines the field-of-view
   of \gem\ \nif . Its Br$\gamma$ images in the velocity ranges of 
[-85, -25], [-25,45] and [35,95] km/s are shown in the lower panel
from left to right, which have the same scale as the image in the upper left.}
\label{f:p112}
 \end{figure*}

\begin{figure*}[!thb]
  \centerline{
      \epsfig{figure=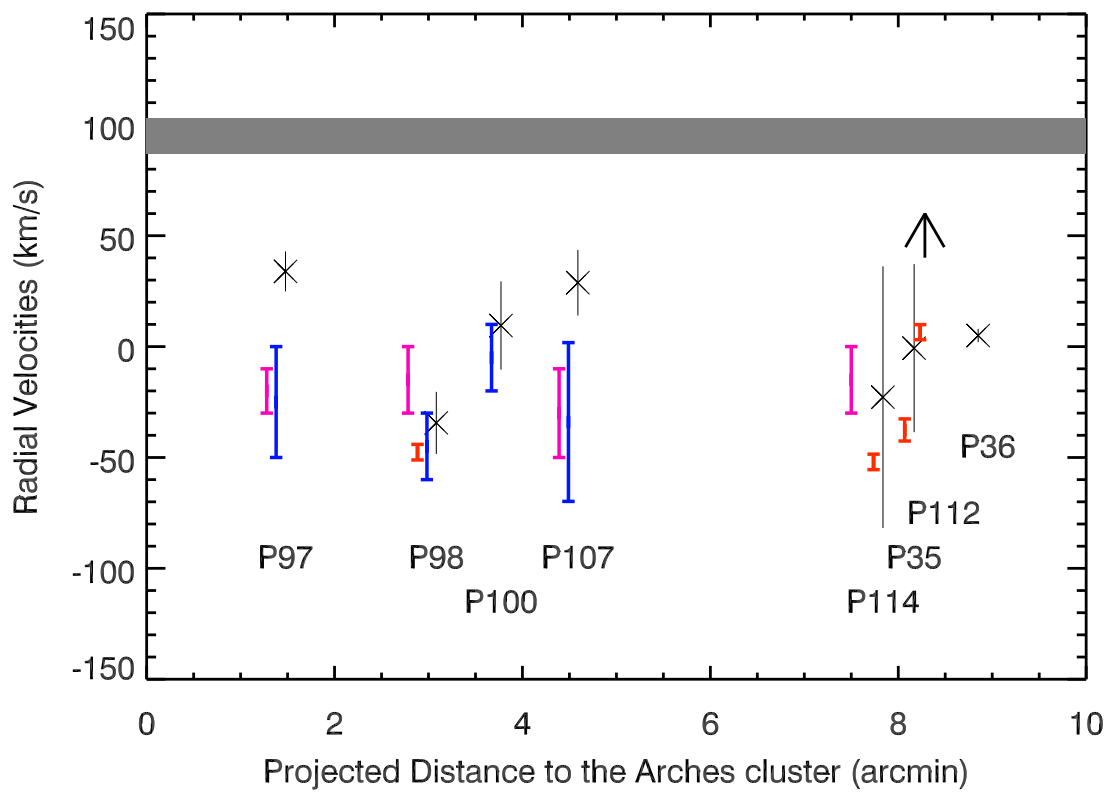,width=1\textwidth,angle=0}
}
 \caption{Comparisons of various radial velocity measurements of our
   program stars (`crosses'; the arrow indicates the lower limit of the
   radial velocity of P112) and their adjacent ionized (Red: this work
   and Blue:~\citealt{lan01b}) and molecular
   (Cyan:~\citealt{ser87,tsu99}) gases. The grey region represents the
   one sigma uncertainty of the radial velocity of the Arches cluster
   from~\citet{fig02}.}
\label{f:los_com}
\end{figure*}

The strongly asymmetric morphologies of the H1/H2 \ion{H}{2} regions are striking (Fig.~\ref{f:source_pos}). The Paschen-$\alpha$ 
emission is the brightest on the side
 towards the -30$-$0 km/s molecular cloud (H1: celestial north, H2:
 celestial northeast, see Fig.~\ref{f:source_pos}). The morphology may
 represent either a bow
shock or a pressure-driven flow in a medium having a strong density gradient~\citep[][and
references therein]{zhu08}. In the bow shock scenario, the stellar
wind from a massive star drives the surrounding gas into a thin
parabolic shell, which captures ionizing photons and generates the observed hydrogen
recombination lines. In the pressure-driven 
flow scenario, the stars are nearly static, relative to the unperturbed surrounding
medium and the flow (commonly called `Champagne flow') is driven by 
the pressure gradient produced by the ionization.

\begin{figure*}[!thb]
  \centerline{
      \epsfig{figure=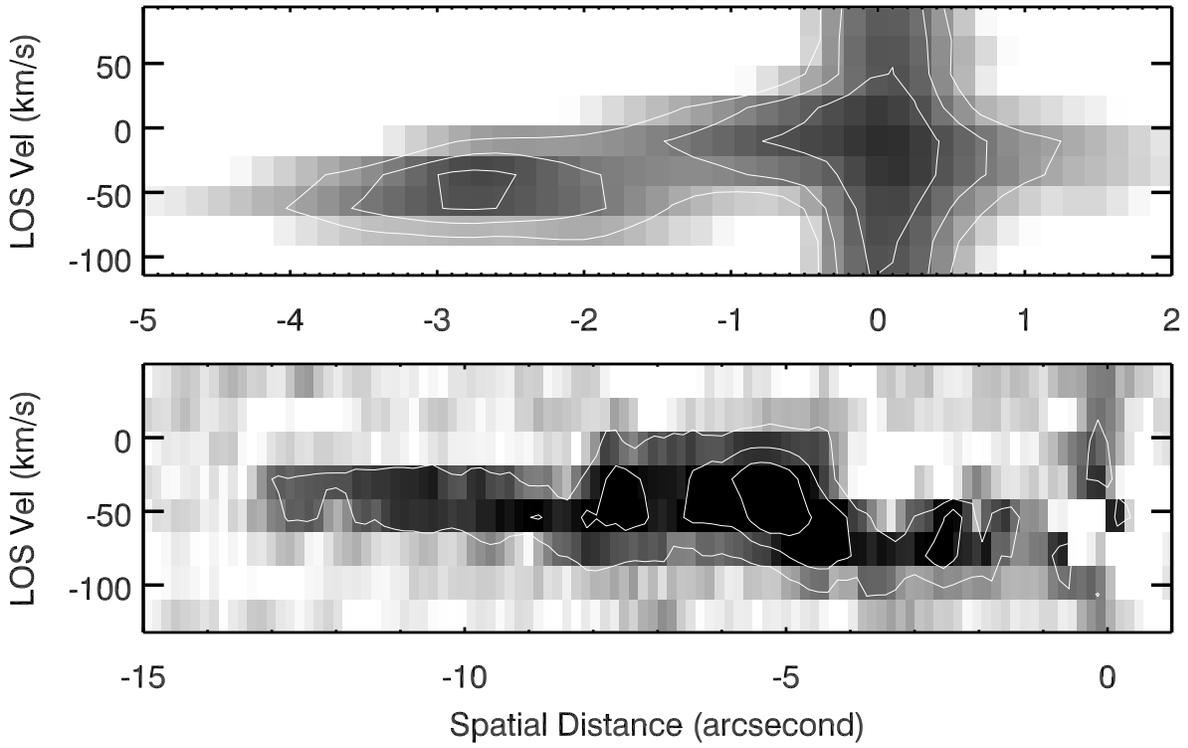,width=1\textwidth,angle=0}
}
 \caption{ Zoomed-in figures of Fig.~\ref{f:p35_spec}c (P35, top) and 
Fig.~\ref{f:p114_spec}c (P114, bottom) between the vertex of the \ion{H}{2} 
regions and the central massive stars. }
\label{f:p35_p114}
\end{figure*}

These two scenarios may be distinguished by their 
different radial velocity distributions of the ionized gas relative 
to the star. A bow shock is expected when a massive 
star moves supersonically through its ambient medium. 
If the motion is not limited to the plane of the sky, 
then the radial velocity of the shock should be about 
the same as that of the star, but should be offset 
from that of the ambient medium. We find that 
the line emission from the bright arcs of H1/H2 is blueshifted, by
 $\sim$40 km/s, relative to
the -30$-$0 km/s molecular cloud. This blue shift suggests that 
the massive stars should be behind the
molecular gas and are moving towards us, relative to the molecular
cloud. According to Fig. 
46 of~\citet{zhu08}, in this situation, when moving from the 
vertex of the shell to massive stars, the radial velocity of the 
ionized gas gets more negative. One
good example is the position-velocity diagram of Sgr A East A derived
from the [\ion{Ne}{2}] 12.8 $\mu$m lines
in~\citet{yus10}, which is used to argue for a bow-shock
model. Similarly, the position-velocity diagram of the ionized gas near P114 in
Fig.~\ref{f:p114_spec}c and Fig.~\ref{f:p35_p114} 
seems to be consistent with that predicted by the
bow shock model. 


The bow shock scenario would require a particular  
geometry in light of the measured extinction of P114, 
which is similar to that of the other
stars (see Table~\ref{t:sample}). Also, the bright 
Paschen-$\alpha$ emission 
from H1 suggests that it arises from the near side 
of the -30$-$0 km/s molecular
cloud. Thus, the bow shock scenario requires that P114 is very close 
to the front edge of and will soon pass through 
the -30$-$0 km/s molecular cloud.~\citet{yus10} invoke a 
similar geometrical juxtaposition to explain the 
Sgr A East A-C \ion{H}{2} regions, which also 
have moderate extinction.  They suggest that the stars
embedded in the Sgr A East A-C regions formed outside and are probably just 
entering a region with low
surface density of molecular gas. Therefore, we cannot 
ignore the possibility that P114
may be an interloper from elsewhere.  More spectroscopic observations
at high velocity resolution are needed to map the H1 
\ion{H}{2} region to explore this possibility. 

If the \ion{H}{2} regions are instead pressure-driven shells, they
would be primarily created by the advance of the ionization front 
into the ambient medium, coupled with the compression by the 
stellar wind. In this scenario, the blueshifted
ionized gas at the vertex of the shell would indicate that the central star 
is in front of the molecular clouds.  This scenario naturally explains the 
moderate extinction of P35.  Unlike the bow shock model, the radial
velocity of the ionized gas approaches that of the central star 
from the vertex of the shell to the central star, which is seen in
Fig 47. of~\citet{zhu08}. A good example of a region where this has happened is the 
G29.96-0.02 \ion{H}{2} region~\citep{zhu08}, which has a
position-velocity diagram along the symmetry axis similar to that we have found for
P35 in Fig.~\ref{f:p35_spec}c and Fig.~\ref{f:p35_p114}.

\subsection{Origins of the Stars}\label{ss:ori}
Where were the massive stars discussed here born? 
Although they were selected as probable GC field stars, they 
could still have formed in clusters, existing
or dissolved. In general, massive stars tend to form in large
aggregates. In the solar neighborhood, for example, 
$>$90\% of the O stars are in, or can be
tracked back to, nearby clusters or OB associations
~\citep[][and references therein]{zin07}. Stars can escape 
from the clusters through supernova explosions 
in binaries, three-body interactions and the strong
tidal forces in the GC.

Our program stars, especially the ones associated 
with nearby \ion{H}{2} regions, could have formed {\sl in-situ}. 
For example, 
one potential site of star formation, G-0.02-0.07, is
$\sim$2.5\arcmin\ (6 pc in projection)
away from Sgr A* and includes four compact \ion{H}{2} 
regions, Sgr A East A-D~\citep{yus10,mil11}. 



\subsubsection{The Three Massive Star Clusters}\label{sss:three}
The three massive star clusters in the GC are the main suspects for
having produced  
these `field' evolved massive stars, because of their unique properties. 
Due to the
high stellar number densities in the cores of the clusters, after the formation of the first
massive binary, three-body interactions among massive 
stars could happen with a large probability~\citep{por10}. The stars with the lowest 
mass are the most likely to be ejected from
the system, with the remaining two more massive 
ones forming a binary system. 
For example, five stars in the Arches cluster 
have initial masses of nearly 120 $M_{\odot}$~\citep{mar08}, larger than all of
our program stars. Three of them have
X-ray counterparts~\citep{wan06} and should be massive binaries. 
A supernova explosion in a binary system could also provide one
massive component of the binary system with enough kinetic energy to
escape the potential well of its cluster. However, this mechanism is
only efficient in clusters that are $>$3 Myrs old, when massive
stars end their lives as supernovae~\citep{gva10}, and is not a very
efficient mechanism to produce young massive runaway stars with escape
velocities $>$ 70 km/s~\citep{gva11}. 

Indeed, \citet{mau10c} find three WNh stars 50\arcsec\ (2 pc, in
projection) off the Arches cluster. These stars have 
magnitudes and spectral types similar to those of the brightest ones in the Arches
cluster and might be coeval 
with the Arches cluster. If they were ejected from the Arches cluster,
soon after they formed, only $<$ 1 km/s proper motions are needed to reach 
their current projected positions. Compared to the three WNh stars, 
our program stars are further away from the Arches cluster,
but have less evolved stellar types. This is consistent
with what one might expect from three-body
interactions: the less massive stars, which evolve more 
slowly, can get higher escape velocities. However, the three-body
interaction mechanism predicts that the stars ejected from the parental
cluster should have random directions, while all our program stars
are blueshifted relative to the Arches cluster. This 
might be a selection effect, however, since the stars included in our observational sample
were selected only to the Galactic west of the Arches cluster.


\citet{gva11} have simulated the three-body interaction process for
massive stars. In one set of their simulations, a star having 
mass $<$ 80 M$_{\odot}$ approaches a binary consisting of 
120 M$_{\odot}$ and 90 M$_{\odot}$ stars at a separation of 0.3 AU. 
As a result of the interaction, the lower-mass single star is 
typically ejected at relatively high velocity, while the binary 
separation decreases.~\citet{gva11} run the
simulation many times with different parameters, and find that the 
escape velocity of the third
star and its mass are anti-correlated. The mean ejection velocity is
$\sim$100 km/s at 20 M$_{\odot}$ and decreases to $\sim$30 km/s at
80 M$_{\odot}$ (Fig.9 of~\citealt{gva11}). They also find that
approximately 1\% of stars with 20/50/80M$_{\odot}$ are expected to
have velocities exceeding 500/300/200 km/s. 

Besides the three-body interaction, massive stars could leave the three
clusters by tidal stripping, particularly when the clusters make their 
closest approach to Sgr A*.~\citet{hab14} simulate the
evolution of the Arches and Quintuplet clusters in the potential well
of the GC, based on their present spatial locations
and 3D velocities. They find that due to the strong tidal field, the tidal
tails of these star clusters could extend to tens of pc. Our program 
stars indeed fall into the Galactic western tidal tail of the Arches
cluster in their simulation. ~\citet{hab14} suggest that their
best-matching model could explain 80\% of the WR stars out to 21 pc
from the Arches cluster and that the velocities of massive stars in the
tidal tail relative to the Arches cluster could reach 20 km/s. 

In Table~\ref{t:vel}, we give the velocities of our program stars,
relative to the Arches cluster. Since we do not have proper motion measurements for our program
stars necessary to derive their full 3D paths, we first consider the
simplest case in which all
eight stars escaped from the Arches cluster 3 Myr ago, the age of the
Arches cluster. If the stars were ejected later in the evolution of the
Arches cluster, 
their proper motion should be higher than what we derive. 
The error of
the required proper motion derived here accounts for the uncertainty in the age of the
Arches cluster (from 2 Myr to 4 Myr), but does not address the
possibility of later
ejections or non-related objects. Combined with the radial
velocities derived in \S~\ref{ss:vel_s}, 
the total velocities of our program stars, relative to the Arches cluster, 
range from 50 to 140 km/s, which are larger than predicted by 
tidal stripping, but still within the velocity
range predicted by the simulations of three-body
interactions by~\citet{gva11}. Therefore, it is possible that all our
program stars were 
previous members of the Arches cluster that left the cluster because
of 
the combined effects of three-body interaction and tidal stripping.

Our new spectra and the stellar ages and masses derived from the NIR photometry
in \S~\ref{s:age} can also help us identify the 
family members of the three clusters. 
The ages of our program stars are
roughly 2-3.5 Myr (see Table~\ref{t:sample}), similar to the Arches
cluster, but younger than the Quintuplet ($\sim$4 Myr old) and the
Central cluster ($\sim$6 Myr old). The four O If$^+$ stars 
(P97, P98, P100 and P107) are the 
closest to the Arches cluster in our sample. Their spectra are
 similar to those of the two O If$^+$ stars, F10 and F15, in
the Arches cluster~\citep[$\sim$2-4 Myr, $\sim$60-90
M$_{\odot}$][]{mar08}. Therefore, they could be previous
members of the Arches cluster. \textbf{P114 has mass}   
similar to those of the OB supergiants in the Arches cluster. 
Although~\citet{osk13} suggest that P114 could have been ejected from
the Central cluster, considering the morphology of H1, the possibility
that P114 was ejected from Arches cluster cannot be ruled out. P114 could lag 
behind the rapidly moving Arches cluster, but still move toward Galactic east,
relative to the old field stars in the GC. We find that no stars in
the Arches cluster have spectra similar to \textbf{P112 and P36}, which might thus have
formed elsewhere. 

Combined with the three WNh stars suggested in~\citet{mau10c}, there
are seven potential `field' evolved massive stars that could be previous members of
the Arches cluster. These stars do not have X-ray counterparts and
seem to be single stars. Following three-body interactions, one should
find the same 
number of massive binaries moving in the opposite directions, 
with velocities less than half those of the ejected single stars, 
relative to the Arches cluster, due to momentum
conservation. If the 
velocities of these massive binaries are small, they cannot
escape the potential well of the Arches cluster. 
Classifying a massive binary system as a
single star would overestimate/underestimate not only its mass/age,
but also those of the whole cluster. 
However, to recognize these close massive binaries in the GC 
is challenging with current near-IR cameras. While the
collision wind zones in massive binaries could emit X-rays, their X-ray luminosities are very 
sensitive to the strength of the stellar winds and the distance
between these two stars. Therefore, in the Arches cluster, 
besides the three potential massive binary systems recognized as X-ray
pointing sources in~\citet{wan06}, we cannot exclude the
possibility that there are additional massive binary systems, which
are X-ray faint. 
A future long-period monitoring
program of the variability and the radial velocity 
of stars in the Arches cluster is needed to explore this possibility
further. Searching for runaway stars in
the other directions from the Arches cluster could also help us
indirectly estimate the number of massive binaries within the cluster. 

\subsubsection{Dissolved Star Clusters}
Besides the three massive clusters, the current field stars could also come from 
dissolved star clusters.~\citet{por02} (see
also~\citealt{kim99,kim00,kim03}) 
simulate the infall of young, massive and compact
star clusters, into the
central potential well. Due to the strong tidal
forces and possible three-body interactions, they find that massive stars can be stripped from
their nascent star clusters.  However, the number densities of 
these clusters are not expected to be below the
background density until after roughly 5 Myr, and are not totally 
dissolved until $\sim$50 Myr. 
As we found in \S~\ref{s:age}, however, the
ages of our observed stars range from 2.2$-$3.2 Myr (see Fig.~\ref{f:iso} and 
Table~\ref{t:sample}). Therefore, they unlikely arise from  
dissolved massive star clusters.


\subsubsection{In-situ Star Formation}\label{sss:insitu}
The massive field stars could also
form in isolation, beyond the star clusters. For example, in \S~\ref{ss:int}, we 
discussed two models to explain the
cometary morphology of the H1/H2 \ion{H}{2} regions, 
which indicate distinct origins for
P114 and P35. In the bow shock scenario, the central star 
is a massive star that formed outside and 
encounters the -30$-$0 km/s 
 molecular cloud with supersonic velocity. In the
 pressure-driven flow scenario, the massive star formed {\sl in-situ}. 
We argue that the pressure-driven flow scenario could be
more suitable for explaining the radial velocities of the adjacent ionized
gas in H2, relative to P35 and the -30$-$0 km/s molecular
cloud. Therefore, P35 could represent the tip of an  
underlying star cluster. If so, P35 is apparently the only 
evolved massive star within the cluster and provides a sufficient UV
photons to ionize the \ion{H}{2} regions. Therefore, this 
new star cluster would have to be substantially smaller
than the known three massive clusters.

How such a small star cluster could form has very important impact on our
understanding of mode of star formation in the GC. The harsh
environment in the GC is very hostile to star
formation~\citep{mor96a}. For example,~\citet{lon13a} find that 
the star formation rate per unit mass of dense gas is an order 
magnitude smaller than that in the Galactic Disk (see also~\citealt{kru14}). 
Due to the high gas temperature/density, strong 
magnetic field and tidal force, only dense molecular clouds could
survive and collapse to form stars, while less dense ones will be tidally 
destroyed. Therefore, the Jeans Mass is very
high~\citep{mor96a} and we should expect the
self-collapse of natal molecular clouds to 
produce the clusters like the three massive one, 
but not a small cluster in H2, especially 
when this star cluster is near the edge, not in the core of the -30$-$0 km/s molecular
clouds. Additional mechanisms, such as external forces, are 
needed to induce collapse. For example, Sgr A East supernova remnant 
was previously suggested to trigger the star formation in Sgr A East
A-D, although later studies suggest that those \ion{H}{2} regions 
are older than Sgr A East supernova remnant~\citep[][and references
therein]{mil11}. Compared to Sgr A East A-D, H2 is a better place to
study the effect of external forces on the star formation in the GC,
because the massive stars ionizing Sgr A East A-D were suggested to be 
interlopers and their birth places are still unclear~\citep{yus10}.  

In order to identify which external forces triggered the star formation
in H2, we need to trace the -30$-$0 km/s molecular
cloud back to its location at 2$-$2.5 Myr ago 
(the age of P35, see Table~\ref{t:sample}).
~\citet{mol11} study the {\sl Herschel} observations of dust emission in
the CMZ and propose that the molecular clouds in the CMZ lie on a 100 pc
twisted ring between Sgr B and Sgr C, with an orbital period of 
$\sim$3 Myr. In their model, interestingly, Sgr A* is
not at the center of this elliptical ring and is close to its front
side, where the known 20 km/s and 50 km/s molecular clouds are
located. On the other hand, the -30$-$0 km/s 
molecular cloud is currently on the back side of the ring. But, 
2$-$2.5 Myr ago the cloud should 
have just encountered Sgr A*.

\citet{lon13a} suggest that Sgr A* could trigger star formation in
the approaching molecular clouds, through compressing them along the
direction perpendicular to their orbits, but stretching in the
direction of the orbit. They argue that the intense star
formation in Sgr B2 could have been triggered by Sgr A*, and the same for four 
dense molecular clouds between Sgr A*
and Sgr B2 above the Galactic Plane. According to the toy model
of~\citet{mol11}, these clouds lie behind Sgr B2 in the twisted
ring and have just passed Sgr A*, which explains why these clouds 
show little sign of early stages of star formation, 
while Sgr B2 is experiencing intense star formation
now. Therefore, the star formation at the edge of the 
-30$-$0 km/s molecular cloud could have been triggered by tidal forces, when it
approached Sgr A* 2$-$2.5 Myr ago. From the {\sl Herschel} observations,
its column density is substantially smaller than those of Sgr B2,
which could explain why any star cluster formed in H2 is 
significantly smaller than that in Sgr B2. 




\section{Summary}\label{s:sum}
In this work, we have measured the radial velocities of eight O
supergiants/O If$^+$ stars in the Galactic center and explored 
their relationship with nearby \ion{H}{2} regions, based primarily on our 
\gem\ \gni/\nif\ spectroscopic observations. We have also constrained 
the ages and masses of these stars from the extinction-corrected 
near-infrared color magnitude diagram. We have further explored
 the origins of these massive stars. Our results are as follows: 

\begin{itemize}
\item We have identified a new O If$^+$ star, P97, $\sim$1.5\arcmin\
  (i.e. 3.6 pc) off the Arches cluster and have refined the spectral
  type of P107 from B[e] to O If$^+$ \textbf{and P36 from O supergiant to O If$^+$.} 

\item We find that the stars
 are 2-3.5 Myr old and have zero-age main-sequence masses $>$50 M$_{\odot}$. 

\item The derived radial velocities of the program stars are all smaller than
  that of the Arches and Quintuplet clusters by at least 50 km/s. We have also
 derived the radial velocities of nearby \ion{H}{2} 
  regions for four of our stars.

\item The radial velocities of P35 in the H2 \ion{H}{2} 
region and the nearby -30$-$0 km/s molecular cloud are similar, but are
  redshifted compared to the ionized gas at the vertex 
  of the shell. The radial velocity pattern of the ionized
  gas along the \gni\ slit suggests that the 
pressure-driven flow scenario is
preferred to explain the cometary morphology of the H2 \ion{H}{2} 
region. Therefore, we find that P35 likely formed {\sl in-situ}. The star
formation could have been triggered 2$-$2.5 Myr ago, perhaps 
when the -30$-$0 km/s
  molecular cloud was near its closest passage to Sgr A*. 

\item The other seven sources are likely unrelated to nearby molecular
  clouds. The bow shock
  scenario can explain the position-velocity diagram of ionized gas near
  P114, which could merely be passing by the -30$-$0 km/s molecular cloud. The
  radial velocities of P97, P107 and P112 are different from nearby
  ionized/molecular gases, suggesting that they should be interlopers. Although the
  radial velocities of P98, P100 and P36 are similar to those of
  nearby \ion{H}{2} regions, their morphologies do not suggest 
that these stars formed {\sl in-situ}. 

\item P97, P98, P100 and P107 have similar ages, masses and spectra to those of
  the two O If$^+$ in the Arches cluster and could then be ejected members. 

\item The origins of P36, P114 and
  P112 remain uncertain. Because of their small ages, they are unlikely from a 
  dissolved massive star cluster. Future proper-motion information
  is needed to determine their birthplaces. 

\end{itemize}

\section*{Acknowledgments}
We thank the anonymous referee for a thorough, detailed, and
constructurive commentary on our manuscript. 
We are grateful to Fabrice Martins, Francisco Najarro, Fahad
Yusef-Zadeh, and Cornelia Lang for many valuable
comments and discussion. H. D. also
acknowledges NASA support via the grant GO-12055 
 provided by the Space Telescope Science Institute, which
is operated by the Association of Universities for Research in
Astronomy, Inc., under NASA contract NAS 5-26555. H. D. appreciates 
Fuyan Bian's great help in the data reduction.

\appendix
\section{Binary Systems}

In the main text, we have assumed that all of 
our program stars are not in binaries. Here we 
consider how the potential binarity may affect our results. 
Our velocity measurement of a star uses certain absorption/emission lines. Thus
depending on whether one member or both contributes significantly to
such lines, the velocity centroid and/or dispersion may differ from
those intrinsic to the binary. In a binary, if only one
member has certain line, the velocity dispersion of this
line in the observed spectrum does not increase 
and the radial velocity derived from its centroid is not equal to
that of the binary. On
the other hand, if both members have the same line, the line in the
observed spectrum will be
broadened.

We may identify massive close 
binaries through their X-ray emission, which comes from
the stellar winds of single massive stars through turbulence-induced
micro shocks or wind collision zone between 
two massive stars. The difference is that the X-ray
spectra of the former type is very soft (kT$\sim$0.6 keV) and could be
totally absorbed by strong foreground extinction toward the
GC, while the later one has hard X-ray spectrum (kT$\sim$1-2 keV). 
We can still detect the X-ray photons with energy larger than 4 keV
from this type of binaries in the GC. In our sample, P35, P36 and P114
have X-ray counterparts~\citep{mau09,mau10a,don12}. The X-ray flux of
P35 and P114 is more than three times larger than that of
P36 and are detected at energies above 4.7 keV, 
while P36 is not. Therefore, P35 and P114 are very likely to be
massive close binaries~\citep{mau10a}. In contrast, the X-ray emission 
from P36 may arise from a single O star~\citep{mau10c} or member stars
of a binary with a relatively large separation. The 
other program stars do not have X-ray counterparts and are not considered to be variable stars
by~\citet{gla01}, indicating that they may not be close massive binaries, but either single massive 
stars, massive binaries with a relative large separation or a binaries, which consist  
of one massive star and one low-mass star. In the latter two cases, the 
velocity derived from certain lines still represents the radial velocity of the system.

Table~\ref{t:velocity} shows that the velocity dispersion
of P36 is similar to that of the four O If$^+$ stars
without X-ray counterparts. Therefore, P36 could be a single massive
star, or with a much less massive companion. In contrast, 
the velocity dispersions ($\Delta
V$) of P35 and P114 are over 100 km/s, which are  
larger than those of the other five O If$^+$
stars (their mean and dispersion 
 are 87 km/s and 9 km/s).~\citet{mas98} find that a large fraction of
spectroscopic binaries have mass ratios close to unity. If so, then the
two stars in P35/P114 binary system may have similar masses and 
the radial velocities derived from 
the centroids of the absorption lines in P35 and P114 are equal to
those of the binaries.

\begin{deluxetable}{cccccccccccc}
\rotate
  \tabletypesize{\tiny}
 \tablecolumns{10}
  \tablecaption{Parameters of Our Sample Stars}
  \tablewidth{0pt}
  \tablehead{
  \colhead{Name}&
  \colhead{R.A.}&
  \colhead{Decl.}&
  \colhead{J} &
  \colhead{H} & 
  \colhead{K} & 
  \colhead{F190N} &
  \colhead{$A_K$} &
  \colhead{Age (Myr)} & 
  \colhead{Mass ($M_{\odot}$)} & 
  \colhead{Type} &
  \colhead{Reference} \\
}
\startdata
P35&266.36926&-28.93473& 14.46(0.02)&11.46(0.02)&9.72(0.03)&
10.59(0.02)& 2.46 & 2.5$\pm$0.4 (2.3$\pm$0.4$^a$) & 102$\pm$17 (98$\pm$17$^a$) & 
  O If$^+$ & 1,2,4 \\
P98&266.42197&-28.86325& 14.80(0.02)&11.60(0.01)&9.86(0.01)&
10.61(0.02)& 2.54 & 2.5$\pm$0.4 & 103$\pm$17 & 
  O9-B0 If$^+$ & 5 \\
P114&266.38658&-28.93794& 14.67(0.02)& 12.13(0.01)& 10.72(0.01)&
  11.35(0.02)& 2.06 & 2.3$\pm$0.8 (2.8$\pm$1.0$^a$) & 83$\pm$21 (60$\pm$19$^a$) & O4-6 I & 3,4 \\
P100&266.42634&-28.87976& 14.70(0.02)& 11.67(0.01)& 10.11(0.03)& 
  10.86(0.02)& 2.35 & 3.2$\pm$1.6 & 72$\pm$22 & O4-6 If$^+$ & 5\\                     
P107&266.41391&-28.88919& 14.68(0.02)& 11.75(0.01)& 10.20(0.01)& 
10.90(0.02)& 2.30 & 2.8$\pm$0.6 & 81$\pm$19 & O4-6 If$^+$ & 1\\
P36&266.38126&-28.95466& 15.13(0.02)& 12.68(0.01)& 11.37(0.02)& 
 11.84(0.02)& 1.96 & 3.2$\pm$1.3 & 54$\pm$21 & O4-6 I & 3,5\\
P97&266.44891&-28.84692& 15.23(0.02) & 12.26(0.01) & 10.54(0.01) &
11.53(0.02)& 2.44 & 2.2$\pm$0.4 &109$\pm$10 & O If$^+$ & \\
P112&266.40753&-28.95450& 17.25(0.03) & 13.06(0.02)& 10.7(0.02)& 12.00(0.02)&
3.19 & 2.7$\pm$0.8 & 93$\pm$20 & B[e] & 1 \\
\enddata
\tablecomments{References for the spectral types: (1)~\citet{cot99};
  (2) ~\citet{mun06}; (3)~\citet{mau09}; (4)~\citet{mau10a};
  (5)~\citet{mau10c}. $^a$ Alterative 
  ages and masses derived from 
  assuming that the two components of each binary are identical.}
\label{t:sample}
\end{deluxetable}


\begin{deluxetable}{ccccccc}
\tabletypesize{\tiny}
\tablecolumns{7}
 \tablecaption{Observation Log}
\tablewidth{0pt}
  \tablehead{
  \colhead{Name}&
\colhead{Instrument} &
\colhead{Band} & 
  \colhead{Observation Date} &
    \colhead{Exposure} & 
   \colhead{Median S/N} &
   \colhead{$V_{LSR,c}^a$ (km/s)}\\
}
\startdata
P35 & GNIRS & H & 2012 Apr 24 (14:41:43.6) & 320 & 21 & -33.8 \\ 
P35 & GNIRS & K & 2012 Apr 24 (14:53:41.6) & 200 & 29 & -33.7 \\ 
P98 & GNIRS & H & 2012 Apr 27 (14:52:06.4) & 480 & 22 & -32.8\\
P98 & GNIRS & K & 2012 Apr 27 (15:02:35.9) & 200 & 27 & -31.3\\
P114 & GNIRS & H& 2012 Jun 6 (11:55:07.7) & 320 & 19 & -15.9\\
P114 & GNIRS & K& 2012 Jun 6 (12:08:20.2) & 168 & 15 & -15.9\\
P100&  NIFS & K & 2012 Jul 1 (08:24:49.6) & 320 & 116 & -3.9 \\                  
P107&  NIFS & K & 2012 Jul 1 (09:20:33.1) & 400 & 105 & -3.8\\
P36& NIFS & K & 2012 Jul 1 (09:46:38.1) & 1200 & 120 & -3.7\\
P97& NIFS & K & 2012 Jul 1 (10:59:01.1) & 600 & 100 & -3.6\\
P112 & NIFS & K & 2012 Jul 2 (09:48:35.6) & 600 & 99 & -3.2\\
\enddata
\tablecomments{$^a$ LSR central velocity}
\label{t:obs}
\end{deluxetable}

\begin{deluxetable}{cccc}
\tabletypesize{\small}
\tablecolumns{4}
 \tablecaption{Velocity Measurements of Spectral Lines}
\tablewidth{0pt}
  \tablehead{
  \colhead{Name}&
\colhead{Lines$^a$} &
 \colhead{$V_{LSR}^b$(km/s)} &
 \colhead{$\Delta$V$^c$(km/s)}
}
\startdata
P35 & \ion{He}{2} 10-7, 12-7 & -0.7$\pm$35.3$\pm$2.6 & 122.6$\pm$1.2 \\
P98 &\ion{H}{1} 10-4, 11-4, 12-4, \ion{He}{1} 3Po-3D &
-34.4$\pm$11.5$\pm$2.5 & 92.3$\pm$3.2 \\ 
P114 & \ion{He}{2} 10-7, 12-7 & -22.8$\pm$56.4$\pm$2.6 & 153.9$\pm$1.5\\ 
P112 & \ion{Fe}{2} & 40.1$\pm$1.9$\pm$3.1 & 41.8$\pm$2.2\\
P112 & \ion{H}{1} 7-4 & 19.7$\pm$0.3$\pm$3.0  & 53.7$\pm$0.4\\
P107 & \ion{He}{2} 10-7, \ion{N}{3} (2S-2Po)
& 28.5$\pm$8.8$\pm$2.9 & 70.8$\pm$0.1 \\
P100 & \ion{He}{2} 10-7, \ion{N}{3} (2S-2Po)
& 9.3$\pm$11.9$\pm$2.9 & 84.6$\pm$0.2 \\
P36 & \ion{He}{2} 10-7, \ion{N}{3} (2S-2Po)
&5.0$\pm$6.1$\pm$2.9 & 91.0$\pm$0.2 \\
P97 & \ion{He}{2} 10-7, \ion{N}{3} (2S-2Po) &
32.1$\pm$4.6$\pm$2.9 & 94.0$\pm$0.3 \\
\enddata
\tablecomments{$^a$ lines used for the velocity measurements of
  individual stars. $^b$ the first term of the error measurements for
  the mean velocity of a star relative to the LSR ($V_{LSR}$) 
 represents the statistic uncertainty
 from `{\tt MPFIT}', while the
  second term is the systematic one introduced by the
  arc calibration. $^c$ $\Delta$V is the velocity dispersion.}
\label{t:velocity}
\end{deluxetable}
\begin{deluxetable}{cc}
\tabletypesize{\small}
\tablecolumns{2}
 \tablecaption{Vacuum Wavelengths of Our Used Near-IR Lines}
\tablewidth{0pt}
  \tablehead{
  \colhead{Lines}&
\colhead{$\lambda$ ($\mu$m) } \\
}
\startdata
\ion{H}{1} 7-4 (Br$\gamma$) & 2.166120 \\ 
\ion{H}{1} 10-4 & 1.736685 \\  
\ion{H}{1} 11-4 & 1.681111 \\  
\ion{H}{1} 12-4 & 1.641167 \\  
\ion{He}{1} (3Po-3D) & 1.700704 \\ 
\ion{He}{2} (10-7) & 2.189113 \\ 
\ion{He}{2} (12-7) & 1.692299 \\ 
\ion{N}{3} (2S-2Po) & 2.247100/2.251300 \\ 
\ion{Fe}{2} (z4Fo-c4F) & 2.089380 \\ 
\enddata
\tablecomments{The wavelengths are from http://www.pa.uky.edu/$\sim$peter/atomic/index.html}
\label{t:vacuum}
\end{deluxetable}
\begin{deluxetable}{ccccc}
\tabletypesize{\small}
\tablecolumns{5}
 \tablecaption{Radial Velocities of Paschen-$\alpha$ Features}
\tablewidth{0pt}
  \tablehead{
  \colhead{Name}&
\colhead{Distance$^a$} &
 \colhead{$V_{LSR}^b$(km/s)} &
 \colhead{$\Delta$V$^c$(km/s)} &
 \colhead{$V_{Rel}^d$(km/s)}
 }
\startdata
P35 & -2.85\arcsec\ & -63.5$\pm$0.4$\pm$2.6 & 22.0$\pm$0.2 & -62.8\\
P35 & 20.85\arcsec\ & -78.3$\pm$1.8$\pm$2.6 & 23.6$\pm$2.4 & -77.6\\
P35 & 29.55\arcsec\ & -39.1$\pm$0.6$\pm$2.6 & 12.4$\pm$4.8 & -38.4 \\
P35 & 43.35\arcsec\ & -37.6$\pm$2.3$\pm$2.6 & 22.0$\pm$2.8 & -36.9\\
P98 & 11.7\arcsec\ & -70.1$\pm$0.7$\pm$2.6 & 14.2$\pm$0.5 & -35.7 \\
P98 & 17.7\arcsec\ & -59.2$\pm$1.1$\pm$2.6 & 17.0$\pm$0.9 & -24.8\\
P98 & 29.7\arcsec\ & -47.6$\pm$0.9$\pm$2.6 & 16.0$\pm$1.0 & -13.2\\
P114 & -6.6\arcsec\ & -39.6$\pm$0.7$\pm$2.6 & 22.8$\pm$0.6 & -16.8\\
P114 & -2.7\arcsec\ & -72.7$\pm$1.3$\pm$2.6 & 18.0$\pm$0.8 & -49.9\\
P114 & 27.75\arcsec\ & -52.0$\pm$0.9$\pm$2.6 & 23.1$\pm$1.1 & -29.2 \\
P112 & $\sim$1\arcsec\ & 6.5$\pm$0.4$\pm$3 & 25.1$\pm$0.3 & $>$-33.6\\
\enddata
\tablecomments{$^a$: distances of the features off the stars along the
  \gni\ slit (see Figs.~\ref{f:p35_spec}-\ref{f:p114_spec}). $^b$ and
  $^c$ are the same as those in Table~\ref{t:velocity}. \textbf{$^d$: the radial 
  velocities of the features relative to the stars.}}
\label{t:hii}
\end{deluxetable}
\begin{deluxetable}{cccc}
\tabletypesize{\normalsize}
\tablecolumns{4}
 \tablecaption{Velocities of Individual Stars relative to the Arches
 Cluster}
\tablewidth{0pt}
  \tablehead{
  \colhead{Name}&
\colhead{$V_{pro}^{a}$(km/s)} &
 \colhead{$V_{los}^{b}$(km/s)} &
 \colhead{$V_{tot}^{c}$(km/s)}
 }
\startdata
P35&6.5+/-2.4&98.7+/-38.7&98.9+/-38.8\\
P98&2.5+/-0.9&132.4+/-16.1&132.4+/-16.2\\
P114&6.2+/-2.3&120.8+/-59.5&121.0+/-59.6\\
P100&3.0+/-1.1&88.5+/-21.4&88.6+/-21.5\\
P107&3.7+/-1.4&69.2+/-16.8&69.3+/-16.9\\
P36&7.0+/-2.6&93.0+/-8.5&93.3+/-8.9\\
P97&1.2+/-0.4&64.1+/-12.0&64.1+/-12.0\\
P112&6.6+/-2.5&57.9+/-8.9&58.3+/-9.3\\
\enddata
\tablecomments{\textbf{$^a$: the proper motion. We assume that these eight stars escaped from the 
Arches cluster 3 Myr ago and thus determine the proper motion from their projected distance to the 
Arches cluster (see \S\ref{sss:three}). $^b$: the radial velocity relative to the Arches cluster. $^c$: the total velocity relative 
to the Arches cluster, the combination in quadrature of the 
proper motion and the radial velocity. The total velocities are larger than predicted by the tidal 
stripping from the simulation in~\citet{hab14}, but still within the velocity range predicted by the simulations 
of three-body interactions by~\citet{gva11}. It means that these stars could be 
previous members of the Arches cluster, but left due to the combined effects of three-body interaction and tidal stripping.}}
\label{t:vel}
\end{deluxetable}
\end{document}